\documentclass[11pt]{article}
\usepackage{hyperref}
\usepackage{amssymb}
\usepackage{graphicx}
\usepackage{subcaption}
\usepackage{float}
\usepackage{slashed}
\usepackage{amsmath}
\usepackage{amssymb}
\usepackage{tikz}
\usepackage{fixmath}
\usepackage{tabularx}
\usepackage{breqn}
\usepackage{authblk}
\usepackage{array}
\usepackage{dsfont}
\usepackage{cite}
\usepackage{csquotes}
\usepackage[section]{placeins}
\usepackage[a4paper, total={6.8in, 10in}]{geometry}
\numberwithin{equation}{section}
\newcommand{\be}{\begin{equation}}
\newcommand{\ee}{\end{equation}}
\newcommand{\bea}{\begin{eqnarray}}
\newcommand{\eea}{\end{eqnarray}}
\newcommand{\ba}{\begin{aligned}}
\newcommand{\ea}{\end{aligned}}

\begin{document}
	
\title{Effect of Noncommutative Geometry on Accretion Disks around RGI-Schwarzschild Black Hole}

\author[]{Dilip Kumar\thanks{21phph03@uohyd.ac.in} }

\affil[]{School of Physics, University of Hyderabad, Central University P.O, Hyderabad-500046, Telangana, India.}

\date{}

\maketitle

\begin{abstract}

In this study, we explore the combined effects of quantum gravity induced by non-commutativity and scale-dependent gravitational coupling on the thermal properties of the thin accretion disks around a Schwarzschild black hole. We consider a $\kappa$-deformed Renormalization Group Improved (RGI) Schwarzschild black hole, where the classical Schwarzschild black hole geometry is modified by the $\kappa$-deformation of space-time and the running Newton's coupling constant $G(r)$. Using the modified metric, we derive the geodesic motion of massive particles, the effective potential, and the thermal properties such as the radiated energy flux, luminosity, and the temperature profile of the accretion disk around the $\kappa$-deformed RGI-Schwarzschild black hole. Our study shows that when non-commutativity is combined with the RGI framework, the effects produce a noticeable deviation from the classical Schwarzschild case. In particular, for small values of the deformation parameter, we observe an increase in the peak energy flux and the temperature of the accretion disk. This suggests that quantum gravity corrections enhance the disk's radiative efficiency, especially in the inner regions closer to the black hole. \\ 

\textit{Keywords :} Non-commutativity, renormalization group, Schwarzschild black hole, accretion disk.

 
\end{abstract}

\section{Introduction} 					\label{Introduction}
 
Accretion disks around black holes are among the most luminous and dynamic structures in the universe, which play a crucial role in various high-energy astrophysical phenomena such as quasars, X-ray binaries, and active galactic nuclei. The study of accretion processes on astrophysical objects was first initiated by Hoyle and Lyttleton in 1939 \cite{Hoyle}. These accretion disks, formed by the infall of matter into the gravitational well of a compact object, emit substantial radiation due to the conversion of gravitational energy into heat. In the non-relativistic regime, early studies of spherical accretion of adiabatic gas onto compact astrophysical objects were carried out \cite{Bondi}. Later, this study extended into the framework of general relativity by examining steady, spherically symmetric accretion onto compact objects, which laid the foundation for understanding the critical behaviour of accreting matter near black holes \cite{Michel,Begelman}. After this, the accretion process has been extensively explored for a variety of black hole space-times such as accretion onto moving black holes \cite{Petrich}, cosmological black holes \cite{Mach}, charged black holes \cite{Jamil}, black holes in higher-dimensional settings \cite{John}, and black hole surrounded by exotic fields such as string clouds \cite{Ganguly}, as well as rotating black holes like the Kerr–Newman solution \cite{Babichev}. Several studies have examined how quantum corrections can influence the radiation flux emitted by thin accretion disks \cite{Shu}. The structure and dynamics of such disks have also been explored in various gravitational settings, including those surrounding compact stellar objects like quark stars \cite{Bhattacharyya}, boson stars \cite{Torres}, and fermion stars \cite{Yuan}.

Previous studies have relied on classical gravity to describe black hole accretion, although classical theories fail at high energies due to their non-renormalizability. In recent years, quantum gravity corrections to classical black hole solutions have gained increasing attention, particularly through the framework of the effective quantum field theory of gravity approach in asymptotically safe gravity \cite{Weinberg}. The concept of asymptotic safety proposes that gravity may remain well-defined in the ultraviolet regime through a non-Gaussian fixed point in its renormalization group flow. This approach has been extended to black hole space-times, where quantum gravity corrections affect the surrounding geometry \cite{Percacci}. This method introduces a scale-dependent gravitational coupling $G(r)$, effectively modifying the Schwarzschild geometry at short distances while preserving general relativity at large scales. The resulting RG-improved Schwarzschild black hole solutions reflect essential quantum effects by incorporating a running Newton constant derived from the renormalization group flow equations \cite{Yang}. These quantum corrections slightly reshape the space-time near a black hole, especially close to the event horizon, where gravity is strongest. As a result, the behaviour of matter falling into the black hole, like the motion of gas in the accretion disk will be affected. Some recent works have shown that these changes can influence important thermal properties of the disk, such as critical velocity, temperature, and energy flux, offering potential signatures of quantum gravity effects in astrophysical observations \cite{Zuluaga}.

Quantum gravity corrections using the renormalization group (RG) are one way to extend classical gravity, while non-commutative geometry provides another promising direction. Non-commutative geometry modifies the structure of space-time by introducing a fundamental length scale. Various studies have investigated the effects of non-commutativity on curved space-times \cite{Douglas, Szabo, Doplicher}. In particular, different models of non-commutative space-times have attracted considerable interest as promising frameworks for incorporating quantum gravity effects into classical theories. For instance, the influence of non-commutative space-time on accretion processes has been studied \cite{Gangopadhyay}, where it was shown that the matter accretion is affected by the strength of the non-commutative parameter in the Moyal space-time. Similarly, the study \cite{Harikumar1} analyzed the $\kappa$-deformed space-time and found that the non-commutativity modifies the geodesic equations, introducing a drag-like term that affects particle trajectories.
 
In several approaches to quantum gravity, non-commutative space-time arises naturally, where the usual smooth structure of space-time is replaced by one defined through non-commuting coordinates. Among the earliest and most studied formulations is the Moyal space-time, where the non-commutativity is defined through constant commutation relations between coordinates. However, Moyal-type non-commutativity generally violates Lorentz symmetry, which poses challenges in constructing consistent relativistic theories \cite{Chaichian}. The alternative non-commutative space-time is $\kappa$-deformed space-time, where the coordinates satisfy Lie-algebraic type commutation relations that preserve rotational symmetry but introduce a fundamental length scale. The $\kappa$-deformation preserves a deformed version of Lorentz symmetry through a Hopf algebraic structure, making it a more robust candidate for describing quantum gravitational corrections \cite{Meljanac,Lukierski,Daszkiewicz}. In this framework, the standard coordinate algebra is modified such that the time coordinate does not commute with spatial coordinates as
\begin{equation}   \label{I-1}
[\hat{x}^i,\hat{x}^j]=0,~~~[\hat{x}^0, \hat{x}^i]=ia\hat{x}^i,~~~a=\frac{1}{\kappa}, 
\end{equation}
where $a$ is the deformation parameter with a dimension of length. This fundamental non-commutativity between time and space leads to a minimal measurable length scale, indicating that the structure of space-time is fundamentally altered at very small scales. Such deformation alters the local geometry and influences the dynamics of particles and fields in regions with strong gravitational fields, such as near black holes or within dense astrophysical objects. Numerous studies have examined the role of $\kappa$-deformation in gravitational and cosmological settings through these theoretical developments. For instance, $\kappa$-deformed corrections to Hawking radiation have been derived using Bogoliubov transformations \cite{Harikumar2}, while compact stars have been analyzed through partition functions and generalized uncertainty principles in a non-commutative framework \cite{Harikumar3}. Various investigations have addressed how $\kappa$-deformation affects black hole thermodynamics and geometry \cite{Gupta} and the structure of superdense stars using modified Einstein field equations \cite{Bhanu}. The minimal length scale alters the behaviour of particles in curved space-time as demonstrated in our previous study \cite{Dilip} on geodesic motion around a $\kappa$-deformed Schwarzschild black hole. These studies highlight the rich phenomenology that arises when quantum gravitational effects are introduced through non-commutative geometry. 

Our previous study constructed a $\kappa$-deformed Schwarzschild metric by incorporating first-order corrections in the deformation parameter $a$, reflecting the non-commutative structure of $\kappa$-deformed space-time. Utilizing this modified metric, we derived the corresponding geodesic equations and investigated the motion of test particles in the equatorial plane. The analysis revealed that $\kappa$-deformation introduces distinct modifications to particle trajectories. We derived the $\kappa$-deformed effective potential from the modified Lagrangian. Additionally, we implemented a numerical multi-particle simulation where a large number of particles with initial conditions followed the $\kappa$-deformed geodesics. Compared to the commutative case, where high angular momentum particles tend to follow wide scattering trajectories, the deformed geometry led to more compact orbits and enhanced particle clustering near the black hole for a longer time. These findings underscore the significant impact of non-commutativity on both individual geodesic structure and collective particle dynamics in black hole space-times. These results reinforce the idea that space-time deformation can significantly influence gravitational dynamics and further motivate a detailed analysis of how such effects impact accretion physics in the presence of strong gravity.

While $\kappa$-deformation and the running gravitational coupling have each been extensively explored as individual approaches to incorporate quantum gravity effects into black hole physics, their combined impact remains relatively unexplored. The $\kappa$-deformation introduces a fundamental length scale through non-commutativity in the space-time, while the renormalization group (RG) approach modifies gravity by making Newton’s constant scale-dependent. Studying their interplay is therefore crucial to understanding how quantum gravitational features jointly influence astrophysical processes. In this work, we address this issue by investigating the geodesic motion and thermal properties of thin accretion disks in the background of a Schwarzschild black hole modified by $\kappa$-deformation in the RG-improved gravity. We analyze how these quantum corrections affect particle trajectories, effective potential, innermost stable circular orbit (ISCO), specific angular momentum, energy, and angular velocity. Furthermore, we examine key observational thermal properties such as energy flux, disk temperature, and differential luminosity to show how the combined effects of non-commutativity and running gravitational coupling may manifest in high-energy astrophysical environments.

This paper is organized as follows. Section \ref{k-deformed} reviews the $\kappa$-deformed space-time and develops the mathematical structure that forms the basis for constructing the $\kappa$-deformed Schwarzschild metric. Section \ref{k-deformed RGI-Schwarzschild} introduces the renormalization group induced running gravitational coupling in $\kappa$-deformed space-time and constructs the $\kappa$-deformed RGI-Schwarzschild space-time metric by incorporating non-commutativity and scale dependence. Section \ref{geodesic} presents the analysis of geodesic motion in this modified geometry, including the constraint on RG parameter, effective potential, ISCO radius, specific angular momentum, specific energy, and angular velocity. Section \ref{accretion} investigates the thermal properties of a thin accretion disk by examining the radiated energy flux, differential luminosity, and temperature profiles across modified space-time geometries. Finally, Section \ref{conclusion} offers summary and conclusion of the work.

\section{$\kappa$-deformation and Schwarzschild space-time} \label{k-deformed}

\subsection{Realization of $\kappa$-deformation}

In $\kappa$-deformed space-time, field theories can be constructed either via the star-product formalism \cite{Dimitrijevic,Daszkiewicz} or by using the realization approach, where non-commutative coordinates are expressed in terms of commutative variables \cite{Meljanac,Meljanac1}. We adopt the realization method to derive the $\kappa$-deformed corrections, allowing us to modify the space-time structure while preserving rotational symmetry. The $\kappa$-deformed coordinates $\hat{x}_{\mu}$ can be represented in terms of the usual commutative coordinates $x_{\mu}$ and their derivatives $\partial_{\mu}$ as \cite{Meljanac}
\begin{equation} \label{ksp-1}
\begin{split}
\hat{x}_0=&x_0\psi(A)+iax_j\partial_j\gamma(A),\\
\hat{x}_i=&x_i\varphi(A),
\end{split}
\end{equation}
where the quantity $A=ia\partial_0=ap^{0}$, and $\psi(A)$, $\gamma(A)$, and $\varphi(A)$ defines a specific realization of the deformed algebra. These functions are required to satisfy the normalization conditions as,
\begin{equation} \label{ksp-2}
\psi(0)=1,~\varphi(0)=1.
\end{equation}   
Substituting Eq.(\ref{ksp-1}) in Eq.(\ref{I-1}), we obtain the constraint between the realization functions as,
\begin{equation} \label{ksp-3}
\frac{\varphi'(A)}{\varphi(A)}\psi(A)=\gamma(A)-1,
\end{equation}
where $\varphi^{\prime}=\frac{d\varphi}{dA}$. There are multiple possible choices for these realization functions. Two common realizations for $\psi(A)$ are $\psi(A)=1$ and $\psi(A)=1+2A$ \cite{Meljanac}. In our study, we adopt the choice $\psi(A)=1$ which simplifies the coordinate expressions mentioned in Eq.(\ref{ksp-1}) as,
\begin{equation} \label{ksp-4}
\begin{split}
\hat{x}_0=&x_0+iax_j\partial_j\gamma(A),\\
\hat{x}_i=&x_i\varphi(A),
\end{split}
\end{equation}
and reduces the constraint equation Eq.(\ref{ksp-3}) as,
\begin{equation} \label{ksp-5}
\frac{\varphi'(A)}{\varphi(A)}=\gamma(A)-1.
\end{equation} 
Several functional forms satisfy these conditions, including  $\varphi(A) = e^{-A}, e^{-\frac{A}{2}}, \frac{A}{e^A-1}, 1$, etc. \cite{Meljanac}. For our study, we adopt $\varphi(A)=e^{-A}$, which corresponds to a commonly used exponential realization in $\kappa$-deformed space-time geometry. It is important to note that the realization functions $\varphi(A)$ and $\gamma(A)$ are not independent but are related through Eq.~(\ref{ksp-5}). 
Once a specific form of $\varphi(A)$ is chosen, the function $\gamma(A)$ is determined from this relation. For instance, adopting the exponential realization $\varphi(A) = e^{-A}$ corresponds to $\gamma(A) = 0$, which is the choice used throughout this work. Now, we incorporate the non-commutativity via $\kappa$-deformation in the Schwarzschild black hole space-time geometry.

\subsection{$\kappa$-deformed Schwarzschild space-time}

To construct the $\kappa$-deformed Schwarzschild metric, we begin by obtaining the $\kappa$-deformed metric from a classical commutative background. In this section, we construct the $\kappa$-deformed metric corresponding to a given background space-time. We begin by assuming that the non-commutative coordinates $\hat{x}_{\mu}$ and their conjugate momenta $\hat{P}_{\nu}$ satisfy the generalized commutation relation as,
\begin{equation} \label{ksp-6}
[\hat{x}_{\mu},\hat{P}_{\nu}] = i\hat{g}_{\mu\nu},
\end{equation}
where the right-hand side is interpreted as the metric tensor of the deformed geometry. We introduce an auxiliary set of coordinates $\hat{y}_{\mu}$ which commute with $\hat{x}_{\mu}$ and satisfy all the properties as $\hat{x}_{\mu}$. Using this, we express the $\kappa$-deformed coordinates and momenta in terms of a chosen realization as \cite{Meljanac,Harikumar4,Kovacevic}
\begin{equation} \label{ksp-7}
\hat{x}_{\mu} = x_{\alpha}\varphi^{\alpha}_{\mu} \, , \quad \hat{P}_{\mu} = g_{\alpha\beta}(\hat{y}) p^{\beta}\varphi^{\alpha}_{\mu} \, ,
\end{equation}
where $\varphi^\alpha_{\mu}$ defines the realization, $p^\beta$ are the commuting momenta corresponding to coordinate $x_{\alpha}$, $\hat{P}_{\mu}$ are the $\kappa$-deformed generalized momenta corresponding to the $\kappa$-deformed coordinate $\hat{x}_{\mu}$, and $g_{\alpha\beta}(\hat{y})$ retains the same functional structure as the original commutative metric but with coordinates replaced by $\hat{y}_{\mu}$, where, $\hat{y}_{\mu}$ commutes with $\hat{x}_{\mu}$ i.e.,
\begin{equation} \label{ksp-8}
[\hat{y}_{\mu},\hat{x}_{\nu}]=0.
\end{equation}
Now, by substituting Eq.(\ref{ksp-7}) in Eq.(\ref{I-1}), we obtain,
\begin{equation} \label{ksp-9}
\varphi _0^0=1, \, \varphi _j^i=\delta _j^i e^{-ap^0},  \, \text{others} \, \, \text{zero.} 
\end{equation}
Also, the $\hat{y}_{\mu}$ appearing in Eq.(\ref{ksp-8}) is used to simplify calculations while preserving the physical content of the non-commutative geometry. Thus, the coordinate $\hat{y}_{\mu}$ satisfies the $\kappa$-deformed space-time commutation relation as,
\begin{equation} \label{ksp-10}
[\hat{y}_0,\hat{y}_i]=ia\hat{y}_i,~~~[\hat{y}_i,\hat{y}_j]=0. 
\end{equation}
A realization of $\hat{y}_{\mu}$ in terms of the commutative coordinates and associated momenta is given by
\begin{equation} \label{ksp-11}
\hat{y}_{\mu} = x_{\alpha} \phi_{\mu}^{\alpha}.
\end{equation}
Using Eq.(\ref{ksp-8}), Eq.(\ref{ksp-10}) \& Eq.(\ref{ksp-11}), we obtain (For detailed calculation, see ref.\cite{Harikumar2})
\begin{equation} \label{ksp-12}
\phi_{0}^{0} = 1, \, \, \phi_{i}^{0} = - ap^{i},  \, \, \phi_{0}^{i} = 0, \, \, \phi_{i}^{j} = \delta_{i}^{j}.
\end{equation}
Substituting Eq.(\ref{ksp-12}) in Eq.(\ref{ksp-11}), we obtain,
\begin{equation} \label{ksp-13}
\hat{y}_0=x_0-ax_jp^j,~~
\hat{y}_i=x_i.
\end{equation}
Substituting the $\kappa$-deformed coordinate $\hat{x}_{\mu}$ and their conjugate momenta $\hat{P}_{\mu}$ in Eq.(\ref{ksp-6}), we obtain
\begin{equation} \label{ksp-14}
 [\hat{x}_{\mu},\hat{P}_{\nu}] \equiv i\hat{g}_{\mu\nu}=ig_{\alpha\beta}(\hat{y})\Big(p^{\beta}\frac{\partial \varphi^{\alpha}_{\nu}}{\partial p^{\sigma}}\varphi_{\mu}^{\sigma}+\varphi_{\mu}^{\alpha}\varphi_{\nu}^{\beta}\Big).
\end{equation}
Substituting Eq.(\ref{ksp-9}) in Eq.(\ref{ksp-14}), we obtained the $\kappa$-deformed metric components as,
\begin{equation}  \label{ksp-15}
\begin{aligned}
\hat{g}_{00}&=g_{00}(\hat{y}),\\
\hat{g}_{0i}&=g_{0i}(\hat{y}) e^{-2ap^{0}}-ag_{im}(\hat{y})p^m e^{-ap^{0}},\\ 
\hat{g}_{i0}&=g_{i0}(\hat{y})e^{-ap^{0}},\\
\hat{g}_{ij}&=g_{ij}(\hat{y})e^{-2ap^{0}}.
\end{aligned}
\end{equation}
Thus, we obtain the line element in $\kappa$-deformed space-time as,
\begin{equation}  \label{ksp-16}
\begin{aligned}
d\hat{s}^2&=g_{00}(\hat{y})dx^0dx^0+\Big(g_{0i}(\hat{y})\big(1-ap^0\big)-ag_{im}(\hat{y})p^m\Big)e^{-2ap^{0}}dx^0dx^i\\&+g_{i0}(\hat{y})e^{-2ap^{0}}dx^idx^0+g_{ij}(\hat{y})e^{-4ap^{0}}dx^idx^j.
\end{aligned}
\end{equation}
In this work, we analyze the dynamics of a test particle in the vicinity of a Schwarzschild black hole. As the line element for classical Schwarzschild space-time metric is given by,
\begin{equation} \label{ksp-17}
ds^2 = -f_{0}(r) dt^{2} + \frac{dr^2}{f_{0}(r)} + r^{2}(d\theta^{2} + \sin^{2}\theta d\phi^{2}) ,
\end{equation}
where the metric function $f_{0}(r)$ for classical Schwarzschild space-time is given by,
\begin{equation} \label{ksp-18}
f_{0}(r) = 1 - \frac{2 G M}{r} \, .
\end{equation}
Given that the cross components in the Schwarzschild metric vanishes (i.e., $g_{0i} = 0$). The resulting $\kappa$-deformed Schwarzschild metric simplifies to,   
\begin{equation}  \label{ksp-19}
d\hat{s}^2=g_{00}(\hat{y})dx^0dx^0+g_{ij}(\hat{y})e^{-4ap^0}dx^idx^j. 
\end{equation}
Substituting from Eq.(\ref{ksp-17}) to Eq.(\ref{ksp-19}), we obtain the line element for $\kappa$-deformed Schwarzschild space-time metric as,
\begin{equation} \label{ksp-20}
d\hat{s}^2= - \bigg\{ 1 - \frac{2 G M}{r} \bigg\} dt^{2} \, + \frac{e^{-4 a p^{0}}}{\bigg\{ 1 - \frac{2 G M}{r} \bigg\}} \, dr^{2} \, + \, e^{-4 a p^{0}} \, \bigg[ r^{2} \, d\theta^{2} \, + \, r^{2} \sin^{2}\theta \, d\phi^{2} \bigg] .
\end{equation}
The exponential factor $e^{-4 a p^{0}}$ represents a constant energy-dependent rescaling of the spatial components of the metric, arising from the $\kappa$-deformation of the spacetime algebra. While this rescaling does not alter the causal structure of the Schwarzschild spacetime for a fixed energy scale, it modifies the proper spatial distances and curvature quantities through the multiplicative scaling. As expected, taking the commutative limit $a \to 0$, leads the expression to revert to its classical Schwarzschild space-time metric.

\section{$\kappa$-deformed Renormalization Group Induced-Schwarzschild space-time} \label{k-deformed RGI-Schwarzschild}

\subsection{Renormalization group induced gravitational coupling} 

At low energy scales, gravitational interactions remain well-described by classical general relativity, and quantum corrections are typically negligible. However, for shorter distances and higher energy regimes approaching the Planck scale, the strength of gravitational interactions increases dramatically. This rapid growth eventually causes the breakdown of a conventional effective field theory of gravity. The asymptotic safety scenario, first introduced by Weinberg \cite{Weinberg} and explored in more detail by \cite{Litim,Niedermaier}, overcomes this problem by proposing that gravitational couplings, particularly Newton’s constant, vary with the energy scale. This quantum effect leads to a running Newton constant, also known as running gravitational coupling. In the framework of asymptotic safety, black hole solutions have been explored by going beyond the usual Einstein–Hilbert action and including additional terms that involve higher powers of space-time curvature, such as the squares of the Ricci scalar, Ricci tensor, and Kretschmann scalar along with gravitational couplings that change with energy scale \cite{Falls,Benedetti}. At low energies of the infrared (IR) regime, quantum corrections to classical Schwarzschild black hole space-times can be effectively captured by replacing Newton’s gravitational constant $G_{0}$ with a running coupling by employing Renormalization Group (RG) techniques \cite{Cai}. 

In this approach, one studies the evolution of scale-dependent effective gravitational action $\Gamma_{k}[g_{\mu \nu}]$, which depends on a coarse-graining scale $k$. The scale dependence of the couplings is governed by the functional renormalization group equation. Within the Einstein–Hilbert truncation, where only Newton's coupling $G(k)$ and the cosmological constant $\lambda_{k}$ are retained, analytic solutions for the flow can be obtained \cite{Bonanno}. This leads to the simple expression as
\begin{equation} \label{RG-1}
	G(k) = \frac{G_0}{1 + \omega G_{0} k^{2}} \, ,
\end{equation}
here $G_{0}$ is the classical Newton's gravitational constant, $\omega$ is a parameter that represents the quantum gravity corrections from non-perturbative renormalization group theory \cite{Bohr}. The running Newton's constant $G(k)$, obtained from the non-perturbative RG flow, acquires physical significance only after the renormalization scale $k$ is related to a coordinate-dependent length-scale. This identification implements the RG-improvement by promoting the classical Schwarzschild coupling to its scale-dependent counterpart, thereby introducing quantum corrections through the replacement $G_{0} \to G(r)$. In the present work, we extend this construction by incorporating space-time non-commutativity via $\kappa$-deformation, leading to the $\kappa$-deformed RGI-Schwarzschild metric.

\subsection{$\kappa$-deformed RGI-Schwarzschild space-time} 

To apply the running gravitational coupling to a curved space-time background, we need to relate the RG momentum scale $k$ to a coordinate-dependent quantity. A common choice is to identify the RG scale with the inverse of the proper radial distance \cite{Bonanno},
\begin{equation} \label{RG-2}
	k(r) = \frac{\xi}{d(r)} ,
\end{equation}
where $\xi$ is a dimensionless numerical constant and $d(r)$ is the proper radial distance from the origin to a radial coordinate $r$ in curved space-time, which is given as,
\begin{equation} \label{RG-3}
	d(r) = \int_0^r \sqrt{g_{rr}(\rho)}\, d\rho
\end{equation}
In the present analysis, we adopt the widely used scale-setting as $k(r) \propto 1/d(r)$, which effectively corresponds to identifying the RG scale with the inverse of the proper radial distance in the weak-field limit. This approximation is well justified in asymptotically flat or weakly curved regions, where the proper distance $d(r)$ approaches the coordinate radius $r$, and has been extensively employed in RG-improved black hole studies \cite{Bonanno}. A more precise scale-setting involving $d(r)$ would be required only in regions of strong curvature, such as near the horizon.

Substituting Eq.(\ref{RG-2}) \& Eq.(\ref{RG-3}) in Eq.(\ref{RG-1}), the running gravitational coupling becomes a position-dependent function as,
\begin{equation} \label{RG-4}
	G(r) = \frac{G_0}{1 + \omega G_0 \xi^2 / d(r)^2}.
\end{equation}
In a $\kappa$-deformed space-time, the underlying non-commutative geometry modifies the measurement of distances. In particular, the radial proper distance acquires an overall rescaling that depends on the deformation parameter $a$ and the energy scale $p^0$ as shown in Eq.(\ref{ksp-15}), thus, we obtain
\begin{equation} \label{RG-5}
	\hat{d}(r) = e^{-2 a p^0}\, d(r).
\end{equation}
Consequently, the RG scale identification is modified to,
\begin{equation} \label{RG-6}
	\hat{k}(r) = \frac{\xi}{\hat{d}(r)} = e^{2 a p^0}\, \frac{\xi}{d(r)}.
\end{equation}
Substituting this into the expression for the running gravitational coupling gives,
\begin{equation} \label{RG-7}
	\hat{G}(r) = \frac{G_0}{1 + \omega G_0 \hat{k}(r)^2}
	= \frac{G_0}{1 + \omega G_0 e^{4 a p^0}\, \xi^2 / d(r)^2}.
\end{equation}
At sufficiently large $r$, the proper distance behaves approximately as $d(r)\sim r$, so,
\begin{equation} \label{RG-8}
	\hat{G}(r) \approx \frac{G_0}{1 + \dfrac{~ \omega G_0 \xi^2 e^{4 a p^0}}{r^2}}.
\end{equation}
With the scale-dependent coupling inserted into the lapse function, we obtain the correction to Newton’s gravitational constant known as running gravitational coupling in non-commutative space-time as,
\begin{equation}  \label{RG-9}
	\hat{G}(r) \simeq G_0 \bigg(1 - \dfrac{~\tilde{\omega} G_0 e^{4 a p^0}}{r^2}\bigg),
	\qquad \tilde{\omega} = \omega \xi^2 .
\end{equation}
Unlike in classical general relativity, where G is constant, here the gravitational coupling $\hat{G}(r)$ exhibits a radial dependence governed by the RG flow in the non-commutative space-time geometry. Substituting the running gravitational coupling in Eq.(\ref{ksp-18}), we obtain the metric function $\hat{f}(r)$ for $\kappa$-deformed RGI-Schwarzschild space-time metric as,
\begin{equation}  	\label{RG-10}
	\hat{f}(r) = \, 1 - \frac{2 \, G_{0} \, M }{r}\bigg( 1 - \frac{~ \tilde{\omega} G_{0} e^{4 a p^{0}}}{r^{2}} \bigg) .
\end{equation}
Thus, we obtain the line element for $\kappa$-deformed RGI-Schwarzschild space-time metric in cosmological unit ($G_{0} = c = 1$) as,
\begin{equation} 	\label{RG-11}
	d\hat{s}_{\text{RGI}}^{2} = - \bigg\{ 1 - \frac{2 M }{r}\bigg( 1 - \frac{~ \tilde{\omega} e^{4 a p^{0}}}{r^{2}} \bigg) \bigg\} dt^{2} + \frac{e^{-4 a p^{0}}}{\bigg\{ 1 - \frac{2 M }{r}\bigg( 1 - \frac{~ \tilde{\omega}  e^{4 a p^{0}}}{r^{2}} \bigg) \bigg\}} dr^2 + r^{2}  e^{-4 a p^{0}}(d\theta^{2} + \sin^{2}\theta d\phi^{2}).
\end{equation}
Here, Eq.(\ref{RG-11}) represents the effect of non-commutativity through $\kappa$-deformation on the running gravitational coupling in the RG-improved Schwarzschild metric. The parameter \(p^{0}\) denotes a fixed background energy scale characterizing the spacetime, typically of the order of the black hole mass or the Planck scale, where $\kappa$–deformation effects become relevant. This modified spacetime metric serves as the effective background geometry for our analysis in the subsequent sections.

\section{Geodesics in the $\kappa$-deformed RGI-Schwarzschild space-time} 	\label{geodesic}

In this section, we investigate the motion of test particles in the $\kappa$-deformed RGI-Schwarzschild space-time. The study of geodesics is crucial for understanding the behaviour of particle motion in the vicinity of black holes, particularly for modeling the accretion disks and studying the observational features such as thermodynamical properties of accretion disk. We begin by determining the location of the event horizon and fixing the value of the running parameter $\tilde{\omega}$ that governs the strength of the quantum correction in the running gravitational coupling $\hat{G}(r)$ arising from the $\kappa$-deformed renormalization group. 

\subsection{Horizon structure and constraints on the running parameter $\tilde{\omega}$}

The line element for the $\kappa$-deformed RGI-Schwarzschild space-time metric derived in the previous section in Eq.(\ref{RG-11}) and the metric function $\hat{f}(r)$ contains the quantum corrections as,
\begin{equation} \label{GH-1}
\hat{f}(r) =  1 - \bigg\{ \frac{2 \, M}{r} \bigg( 1 - \frac{~ \tilde{\omega} e^{4 a p^{0}}}{r^{2}} \bigg) \bigg\} ,
\end{equation}	
The radius of the event horizon of the improved metric can be calculated by solving $g_{00} = 0$, i.e., $\hat{f}(r) = 0$, we obtain,	
\begin{equation} \label{GH-2}
r^{3} - 2 M r^{2} + 2 M \tilde{\omega} e^{4 a p^{0}} = 0 .
\end{equation}
Solving the above Eq.(\ref{GH-2}) gives the real root as,
\begin{multline}  \label{GH-3}
r_{1} = \frac{2 M}{3} - \frac{4 M^{2}}{3 \sqrt[3]{- 8 M^{3} + 27 M \tilde{\omega} e^{4 a p^{0}} + \sqrt{-64 M^{6} + M^{2} (- 8 M^{2} + 27 \tilde{\omega} e^{4 a p^{0}})^{2} }}} \\ - \frac{\sqrt[3]{- 8 M^{3} + 27 M \tilde{\omega} e^{4 a p^{0}} + \sqrt{-64 M^{6} + M^{2} (- 8 M^{2} + 27 \tilde{\omega} e^{4 a p^{0}})^{2} }}}{3} .
\end{multline}	
We can check the classical Schwarzschild case ($\tilde{\omega} \to 0 ; a \to 0$), gives the well known result $r_{1} = 2 M $, which is the classical Schwarzschild radius ($r_{S}$).	

We calculate the running parameter $\tilde{\omega}$ for our study of $\kappa$-deformed RGI-Schwarzschild metric by minimizing the real root of radius of event horizon from Eq.(\ref{GH-3}) (as $\frac{d r_{1}}{d \tilde{\omega}} = 0$), and obtain the condition,	
\begin{multline}   \label{GH-4}
\frac{1}{9} \underbrace{\bigg( 27 M e^{4 a p^{0}} + \dfrac{M^{2} (- 8 M^{2} + 27 \tilde{\omega} e^{4 a p^{0}}) 27 e^{4 a p^{0}}}{\sqrt{- 64 M^{6} + M^{2} (- 8 M^{2} + 27 \tilde{\omega} e^{4 a p^{0}})^{2}} } \bigg)}_{\text{I$^{st}$ term}} \\ . \underbrace{\bigg( - 8 M^{3} + 27 M \tilde{\omega} e^{4 a p^{0}} + \sqrt{-64 M^{6} + M^{2}( - 8 M^{2} + 27 \tilde{\omega} e^{4 a p^{0}})^{2} } \bigg)^{- 2/3}}_{\text{II$^{nd}$ term}} \\ . \underbrace{\bigg\{ 4 M^{2} \bigg(- 8 M^{3} + 27 M \tilde{\omega} e^{4 a p^{0}} + \sqrt{-64 M^{6} + M^{2} (- 8 M^{2} + 27 \tilde{\omega} e^{4 a p^{0}})^{2}} \, \bigg)^{- 2/3}  - 1 \bigg\}}_{\text{III$^{rd}$ term}} = 0 .
\end{multline}
Since, the I$^{st}$ and II$^{nd}$ terms are the only possibility of $M = 0$. Thus, we have to consider
\begin{equation} \label{GH-5}
 4 M^{2} \bigg(- 8 M^{3} + 27 M \tilde{\omega} e^{4 a p^{0}} + \sqrt{-64 M^{6} + M^{2} (- 8 M^{2} + 27 \tilde{\omega} e^{4 a p^{0}})^{2}} \, \bigg)^{- 2/3}  - 1 = 0 .
\end{equation}
Solving above Eq.(\ref{GH-5}) by considering the mass of black hole ($M=1$) which is much higher than the mass of particles moving around, we obtain,
\begin{equation} \label{GH-6}
\tilde{\omega}_c = \dfrac{16}{27} ~ e^{- 4 a p^{0}} .
\end{equation}

In the context of $\kappa$-deformed RGI-Schwarzschild metric, our results show that the existence of an event horizon depends critically on the value of the running parameter $\tilde{\omega}$. We find that an event horizon forms only when $\tilde{\omega}$ is less than or equal to the critical value, i.e., $\tilde{\omega} \le \tilde{\omega}_{c} = \frac{16}{27} ~ e^{- 4 a p^{0}}$. If $\tilde{\omega}$ exceeds the critical value, the geometry no longer supports a horizon, indicating a breakdown of the black hole structure, which could lead to a naked singularity. This critical value ensures the consistency of the space-time geometry under quantum gravity corrections.

\begin{figure}[H]
	\centering
	\includegraphics[scale=.55]{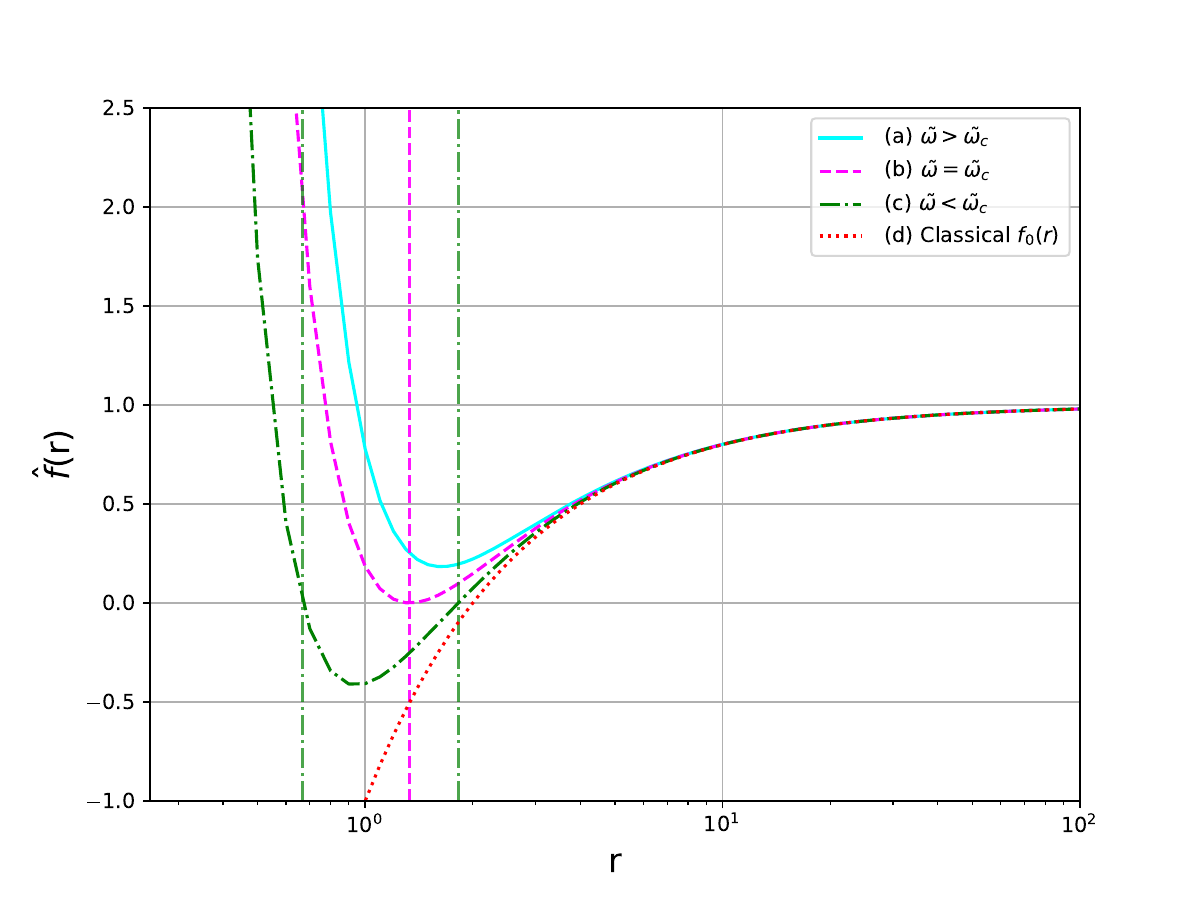}
	\caption{The plot shows the comparison of the improved Schwarzschild metric $\hat{f}(r)$ for three different values of the running parameter $\tilde{\omega}$, (a) $\tilde{\omega} > \tilde{\omega}_{c}$ (solid cyan), (b) $\tilde{\omega} = \tilde{\omega}_{c}$ (dashed magenta), and (c) $\tilde{\omega} < \tilde{\omega}_{c}$ (dashed-dotted green), along with the classical Schwarzschild metric $f_{0}(r)$ (dotted red). The corresponding vertical lines indicate the locations of the event horizon for cases (b) and (c), showing that the horizon vanishes for case (a).}
	\label{Fig1-f(r)vsr}
\end{figure}

In order to analyze the presence of singularities in the $\kappa$-deformed RGI-Schwarzschild geometry, we examine the behaviour of the metric function $f(r)$ near $r=0$. Unlike the classical Schwarzschild case, where $f(r)\rightarrow \infty$ as $r\rightarrow0$, the running of $\hat{G}(r)$ via $\tilde{\omega}$ introduces a suppression of the gravitational interaction at small scales. Thus, we plotted $\hat{f}(r)$ against $r$ for three different cases of $\tilde{\omega}$ as shown in Fig.(\ref{Fig1-f(r)vsr})  

Since the effects of non-commutativity are expected to be small, we obtain $a p^{0} \simeq 0.1$, where $p^{0}$ denotes the energy scale associated with the $\kappa$-deformed space-time, typically taken to be of the order of the black hole mass. At such high energies comparable to the Planck scale, non-commutative corrections become relevant \cite{Bhanu}. From Fig.(\ref{Fig1-f(r)vsr}), we have obtained that for the critical value $\tilde{\omega} \simeq 0.39 = \tilde{\omega}_c$, the two horizons merge, corresponding to a black hole. For $\tilde{\omega} \simeq 0.19 < \tilde{\omega}_c$, the space-time admits two distinct horizons such that one is an inner Cauchy horizon and the other is an outer event horizon. For $\tilde{\omega} \simeq 0.59 > \tilde{\omega}_c $, there is no event horizon, which means a naked singularity develops. This analysis demonstrates how the causal structure of the $\kappa$-deformed RGI-Schwarzschild space-time is sensitively influenced by the value of the running parameter $\tilde{\omega}$. After analyzing the horizon structure and singularity behaviour of the $\kappa$-deformed RGI-Schwarzschild space-time, we study the motion of test particles by explicitly solving the geodesic equations derived from the modified line element.

\subsection{Test particle trajectories in  $\kappa$-deformed RGI-Schwarzschild space-time }   

In this subsection, we analyze the trajectory of a test particle moving around the $\kappa$-deformed RGI-Schwarzschild space-time. We begin with the modified Lagrangian as \cite{Dilip},
\begin{equation} \label{GT-1}
\mathcal{\hat{L}} = \frac{1}{2} \hat{g}_{\mu \nu} \frac{d \hat{x}^{\mu}}{d \tau} \frac{d \hat{x}^{\nu}}{d \tau} ,  
\end{equation} 
from Lagrangian, we derive the geodesic equations as,
\begin{equation} \label{GT-2}
\frac{d^{2} \hat{x}^{\mu}}{d \tau^{2}}  +  \hat{\Gamma}^{\mu}_{\rho \sigma} \frac{d \hat{x}^{\rho}}{d \tau} \frac{d \hat{x}^{\sigma}}{d \tau}= 0 ,  
\end{equation} 
where the Christoffel symbols ($\hat{\Gamma}^{\mu}_{\rho \sigma}$) are given by,
\begin{equation} \label{GT-3}
\hat{\Gamma}^{\mu}_{\rho \sigma}=\frac{1}{2}\hat{g}^{\mu \nu}\Big(\partial_{\sigma}\hat{g}_{\rho \nu}+\partial_{\rho}\hat{g}_{\nu \sigma}-\partial_{\nu}\hat{g}_{\rho \sigma}  \Big) .  
\end{equation} 
Substituting the line element for the $\kappa$-deformed RGI-Schwarzschild black hole from Eq.(\ref{RG-11}) into the geodesic equation above, we compute the relevant components that govern the particle's motion. The dependence on the proper time $(\tau)$ is obtained as, 	
\begin{equation} \label{GT-4}
\frac{d^{2} t}{d \tau^{2}}  + e^{-a p^{0}} \frac{ 2 M (- r^{2} + 3 \tilde{\omega} e^{4a p^{0}})}{r (2 M (r^{2} - \tilde{\omega} e^{4a p^{0}}) - r^{3})} \, . \,  \, \frac{d \, r}{d \tau} \, . \, \frac{d \, t}{d \tau} = 0,
\end{equation}
We visualize the trajectories in three dimensions using spherical polar coordinates $(r, \theta, \phi)$ for massive particles. The radial equation governing the orbital motion is obtained as,  		
\begin{multline} \label{GT-5}
\frac{d^{2} r}{d \tau^{2}}  + e^{5 a p^{0}}~\bigg( \frac{M (r^{2} - 3 \tilde{\omega} e^{4a p^{0}}) (- 2 M (r^{2} - \tilde{\omega} e^{4a p^{0}}) + r^{3} )}{r^{7}} \bigg) \bigg(\frac{d \, t}{d \tau}\bigg)^{2} \\
- e^{-a p^{0}}~ \bigg( \frac{M (r^{2} - 3 \tilde{\omega} e^{4a p^{0}}) (- 2 M (r^{2} - \tilde{\omega} e^{4a p^{0}}) + r^{3} )}{r (2 M (r^{2} - \tilde{\omega} e^{4a p^{0}}) - r^{3})^{2} } \bigg)  \bigg(\frac{d \, r}{d \, \tau}\bigg)^{2} \\
- e^{-a p^{0}}~ \Bigg( \frac{(- 2 M (r^{2} - \tilde{\omega} e^{4a p^{0}}) + r^{3})}{r^{2}} \Bigg) \, \Bigg\{ \Bigg(\frac{d \, \theta}{d \tau} \Bigg)^{2} + \sin^{2}\theta \Bigg(\frac{d \, \phi}{d \, \tau} \Bigg)^{2} \Bigg\} = 0 , 
\end{multline}
Similarly, the equations governing the angular coordinates $\theta$ and $\phi$, obtained as
\begin{equation} \label{GT-6}
\dfrac{d^{2} \theta}{d \tau^{2}}  +   e^{-a p^{0}} \frac{2}{r} \bigg( \dfrac{d \, r}{d \, \tau}\bigg) \bigg(\dfrac{d \, \theta}{d \, \tau} \bigg) - e^{-a p^{0}} \frac{\sin(2\theta)}{2} \bigg(\dfrac{d \, \phi}{d \, \tau} \bigg)^{2} = 0,
\end{equation}
and 		
\begin{equation} \label{GT-7}
\dfrac{d^{2} \phi}{d \tau^{2}}  +   e^{-a p^{0}} \frac{2}{r} \bigg( \dfrac{d \, r}{d \, \tau}\bigg) \bigg(\dfrac{d \, \phi}{d \, \tau} \bigg) + e^{-a p^{0}} \frac{\sin(2\theta)}{\sin^{2}(\theta)} \bigg(\dfrac{d \, \theta}{d \, \tau} \bigg) \bigg(\dfrac{d \, \phi}{d \, \tau} \bigg) = 0. 
\end{equation}

In all the geodesic equations derived above, the deformation parameter $a$ introduces corrections due to the non-commutativity of space-time, while the running parameter $\tilde{\omega}$ reflects the scale-dependent nature of gravity through the renormalization group improvement in the non-commutative space-time geometry. The equations incorporate contributions from both effects, the non-commutativity via $\kappa$-deformation parameter $a$ and the modified gravitational potential arising from the running gravitational coupling $\hat{G}(r)$. In the commutative limit $a \rightarrow 0$ and classical limit $\tilde{\omega} \rightarrow 0$, the geodesic equations reduce to those of the classical Schwarzschild space-time.

\begin{figure}[H]
	\centering
	\includegraphics[scale=.55]{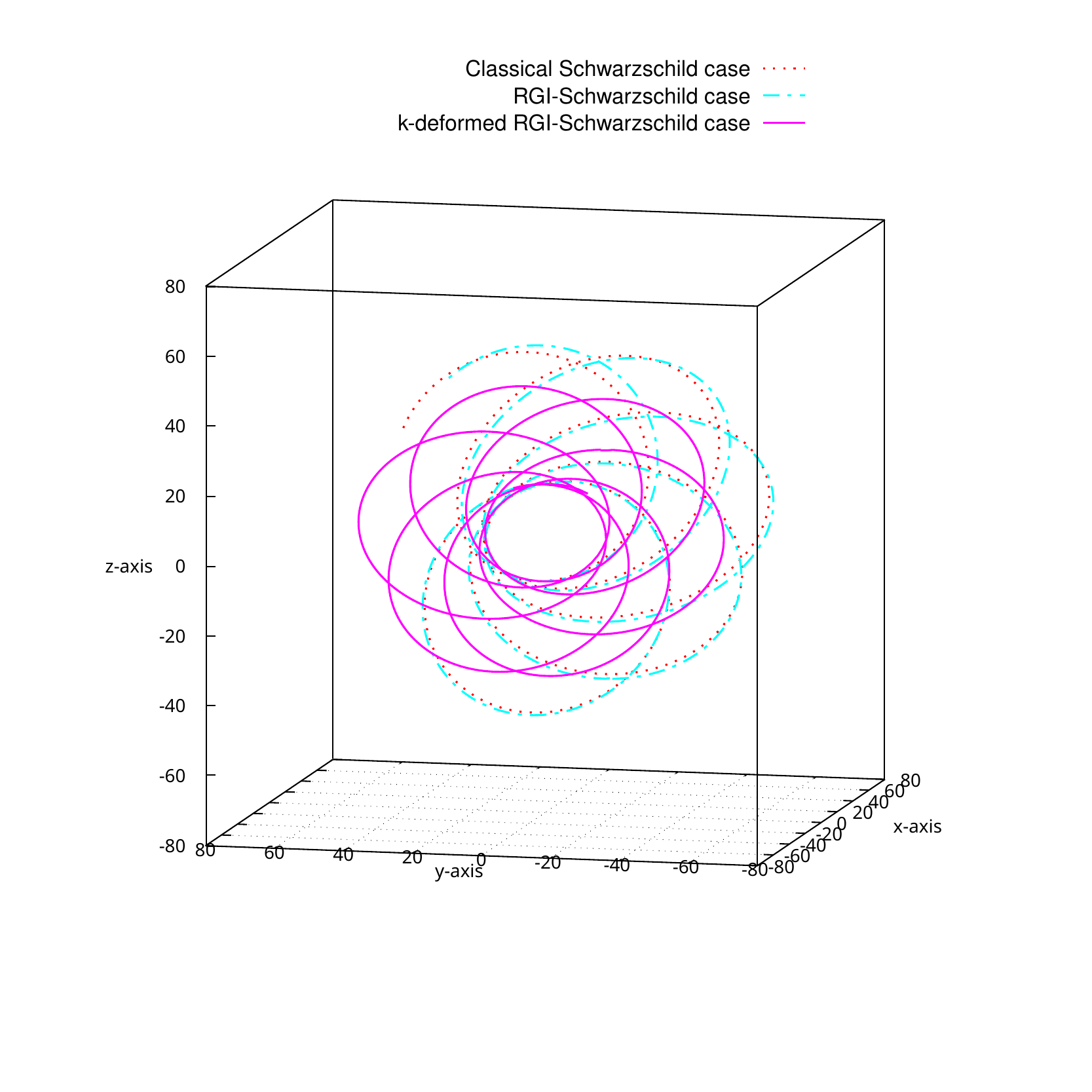}
	\vspace{-2cm}
	\caption{The plot shows the geodesic motion of a massive test particle around the Schwarzschild black hole in different space-time backgrounds. The dotted line (red color) shows the classical Schwarzschild case, the dashed-dotted line (cyan color) corresponds to the RGI-Schwarzschild case, the solid line (magenta color) represents the $\kappa$-deformed RGI-Schwarzschild case.}
	\label{Fig2-GeodesicMotion}
\end{figure}

We numerically solve the geodesic equations for various initial conditions, including positions, angular momenta, and velocities. To incorporate the effect of non-commutativity and renormalization group-induced space-time, we compare the motion of particles under identical initial conditions across three cases: the classical Schwarzschild case ($ap^{0}=0, \, \tilde{\omega}=0$), the RGI-Schwarzschild case ($ap^{0}=0, \, \tilde{\omega}=16/27$), and the $\kappa$-deformed RGI-Schwarzschild case ($ap^{0}=0.1, \, \tilde{\omega}=0.39$). We observe that including the running gravitational coupling and non-commutativity leads to noticeable deviations in the trajectories. In particular, the orbits become more compact and exhibit stronger curvature than the classical case shown in Fig.(\ref{Fig2-GeodesicMotion}).

The particle trajectories shown in Fig.(\ref{Fig2-GeodesicMotion}) demonstrate the effect of non-commutativity and the running gravitational coupling on particle motion around the Schwarzschild black hole. In the classical Schwarzschild case, the particle moves in a wide, smoothly precessing elliptical orbit. In the case of RGI-Schwarzschild case, we observe a deviation from the classical Schwarzschild case. In the $\kappa$-deformed RGI-Schwarzschild case, we observe a difference from both the RGI-Schwarzschild case and the classical Schwarzschild case. The particle’s orbit is much smaller, which shows that gravity is stronger near the $\kappa$-deformed RGI-Schwarzschild black hole. The orbit also precesses more, and the particle takes less time to complete one orbit compared to the classical Schwarzschild and RGI-Schwarzschild case, which means it stays closer to the black hole. These changes would affect how matter behaves in accretion disks and how radiation is emitted from such black holes. These results show that the combine effect of non-commutative geometry and quantum gravitational corrections significantly influence particle dynamics in the strong-field regimes.

\subsection{Effective potential and ISCO radius in $\kappa$-deformed RGI-Schwarzschild space-time}

The effective potential $V_{\text{eff}}$ helps us understand how a particle moves around a black hole. It shows how gravity pulls the particle inward while its angular momentum pushes it outward. By looking at the shape of $V_{\text{eff}}$, we can tell whether the particle will fall into the black hole, escape to infinity, or stay in a stable orbit. In our study, the effective potential in the background of a $\kappa$-deformed RGI-Schwarzschild black hole, where non-commutative geometry (via the deformation parameter $a$) and scale-dependent quantum gravitational correction (via the running parameter $\tilde{\omega}$) modify the space-time structure which becomes essential to study how these corrections influence the effective potential.

The line element in Eq.(\ref{RG-11}) can be rewritten in a form similar to the classical Schwarzschild metric by introducing an effective mass, $\hat{M}_{\text{eff}}(r)$, which depends on the radial coordinate $r$, deformation parameter $a$ and the running parameter $\tilde{\omega}$ as,
\begin{equation} \label{GE-1}
\hat{M}_{\text{eff}}(r) = M \bigg(1 - \frac{~\tilde{\omega} e^{4 a p^{0}}}{r^{2}} \bigg) \, .
\end{equation}
Here, the effective mass $\hat{M}_{\text{eff}}(r)$ formally vanishes at $r^{2} = \tilde{\omega} e^{4 a p^{0}}$. However, this point lies outside the physically relevant domain of the effective description. The effective mass remains positive for $r > \sqrt{\tilde{\omega}}\, e^{2 a p^{0}}$, where the geometry continues to describe a physically meaningful black hole spacetime.

We begin with the Lagrangian for a point particle moving in the $\kappa$-deformed RGI-Schwarzschild space-time. It is given by,
\begin{equation} \label{GE-2}
\mathcal{\hat{L}}_{\text{RGI}} = \frac{1}{2} \hat{g}_{\mu \nu}(a) \dot{\hat{x}}^{\mu} \dot{\hat{x}}^{\nu} = - \frac{1}{2} \delta \, .
\end{equation}
where $ \dot{\hat{x}}^{\mu} = \frac{d\hat{x}^{\mu}}{d\sigma}$ denotes the derivative with respect to the affine parameter $\sigma$, and $\delta = 1, 0$ is for massive and massless particles, respectively. The metric $\hat{g}_{\mu \nu}(a)$ represents the deformation due to the non-commutative geometry. Using the $\kappa$-deformed RGI-Schwarzschild space-time, we now compute the explicit form of the Lagrangian $\mathcal{\hat{L}}_{RGI}$ as,  
\begin{equation} \label{GE-3}
\mathcal{\hat{L}}_{RGI} = - \frac{1}{2} \bigg\{ 1 - \frac{2 \hat{M}_{\text{eff}}(r)}{r}\bigg\} \dot{t}^{2} + \frac{1}{2} \frac{e^{-4 a p^{0}}}{\bigg\{ 1 - \bigg(\frac{2 \, \hat{M}_{\text{eff}}(r)}{r} \bigg) \bigg\}}  \dot{r}^{2} + \frac{1}{2} e^{-4 a p^{0}} r^{2} \dot{\theta}^{2} + \frac{1}{2} e^{-4 a p^{0}} r^{2} \sin^{2}\theta \dot{\phi}^{2} = - \frac{1}{2} \delta \, . 
\end{equation} 
From above Eq.(\ref{GE-3}), we obtain the constant equation of motion for $\mu = t, \phi$ as
\begin{equation} \label{GE-4}
\Bigg( 1 - \frac{2 \hat{M}_{\text{eff}}(r)}{r} \Bigg) \, \dot{t} = - \hat{k} \, \, \text{(constant)},
\end{equation}
and
\begin{equation} \label{GE-5}
r^{2} \, e^{-4 a p^0} \, \dot{\phi} = \hat{h} \, \, \text{(constant)}.
\end{equation}
Here, $\hat{k}$ and $\hat{h}$ are constants. For simplification, we confine the motion to the equatorial plane by setting $\theta = \pi/2$. As a result, the effective potential describes particle trajectories only within this plane. Now, by substituting the value of constant equations of motion from Eq.(\ref{GE-4}) \& Eq.(\ref{GE-5}) in Eq.(\ref{GE-3}), we obtain,
\begin{equation} \label{GE-6}
\frac{1}{2} k \dot{t} + \frac{1}{2} \frac{e^{-4 a p^{0}}}{\bigg\{ 1 - \bigg(\frac{2 \, \hat{M}_{\text{eff}}(r)}{r} \bigg) \bigg\}}  \dot{r}^{2} + 0 + \frac{1}{2} \, \hat{h} \, \dot{\phi} = - \frac{1}{2} \delta . 
\end{equation}  
By solving the Eq.(\ref{GE-6}), we obtain,
\begin{equation} \label{GE-7}
\frac{1}{2} \, \dot{r}^{2} - \frac{\delta}{r} \, \hat{M}_{\text{eff}}(r) \, e^{4 a p^{0}} + \frac{\hat{h}^{2}}{2 r^{2}} \, e^{8 a p^{0}} - \frac{\hat{h}^{2}}{r^{3}} \, \hat{M}_{\text{eff}}(r) \, e^{8 a p^{0}} =  \frac{1}{2} \, e^{4 a p^{0}} \, (k^{2} - \delta). 
\end{equation}
We compare the above Eq.(\ref{GE-7}) with the energy equation which is given by,
\begin{equation} \label{GE-8}
\frac{1}{2} \dot{r}^{2} + V_{\text{eff}}(r) =  \text{Constant} \, . 
\end{equation}
Thus, we obtain the effective potential for the $\kappa$-deformed RGI-Schwarzschild black hole as,
\begin{equation} \label{GE-9}
\hat{V}_{\text{eff}}^{\text{RGI}}(r) =  - \frac{\delta}{r} \hat{M}_{\text{eff}}(r) \, e^{4 a p^{0}} + \frac{\hat{h}^{2}}{2 r^{2}} e^{8 a p^{0}} - \frac{\hat{h}^{2}}{r^{3}} \hat{M}_{\text{eff}}(r) \, e^{8 a p^{0}}. 
\end{equation}
To determine the particle orbit around the black hole, we aim to express $r$ as a function of $\phi$ \cite{Hobson}. Thus, we do the change in variable as $u = \frac{1}{r}$, $\implies \frac{\partial u}{\partial \phi} = - \frac{1}{r^{2}} \frac{\partial r}{\partial \phi}$, and $\dot{r} = \frac{\partial r}{\partial \tau} =  \frac{\partial r}{\partial \phi}  \frac{\partial \phi}{\partial \tau}$. Now, by putting the value of $\dot{\phi}$ from Eq.(\ref{GE-5}), we obtain,
\begin{equation} \label{GE-10}
\dot{r} = - \frac{\hat{h}}{e^{- 4 a p^{0}}} \frac{\partial u}{\partial \phi} . 
\end{equation}
Substituting the value of $\dot{r}$ from Eq.(\ref{GE-10}) in Eq.(\ref{GE-7}), we obtain,
\begin{equation} \label{GE-11}
\frac{1}{2} \hat{h}^{2} e^{8 a p^{0}} \bigg(\frac{\partial u}{\partial \phi}\bigg)^{2} - \delta \, u \, \hat{M}_{\text{eff}}(u) \, e^{4 a p^{0}} + \frac{\hat{h}^{2}}{2} u^{2} e^{8 a p^{0}} - \hat{h}^{2} u^{3} \hat{M}_{\text{eff}}(u) \, e^{8 a p^{0}} =  \frac{1}{2} e^{4 a p^{0}} (k^{2} - \delta). 
\end{equation} 
Simplifying the above equation and we obtain as,
\begin{equation} \label{GE-12}
\bigg(\frac{\partial u}{\partial \phi}\bigg)^{2} + u^{2} =  \frac{k^{2} - \delta}{\hat{h}^{2} e^{4 a p^{0}}} + \frac{2 \, u \, \delta \, \hat{M}_{\text{eff}}(u)}{\hat{h}^{2} e^{4 a p^{0}}} + 2 \, u^{3} \, \hat{M}_{\text{eff}}(u) . 
\end{equation}
Substituting value of $\hat{M}_{\text{eff}}(u)$ from Eq.(\ref{GE-1}) as $\hat{M}_{\text{eff}}(u) = M (1 - \tilde{\omega} e^{4a p^{0}} u^{2})$ and differentiating above equation with respect to $\phi$, we obtain as,
\begin{equation} \label{GE-13}
\frac{\partial^{2} u}{\partial \phi^{2}} + u =  \frac{\delta \, M}{\hat{h}^{2} e^{4 a p^{0}}} \bigg(1 - 3 \, \tilde{\omega} e^{4a p^{0}} \, u^{2} \bigg) +  M \, u^{2} \, ( 3 - 5 \, \tilde{\omega} e^{4a p^{0}} \, u^{2}) . 
\end{equation}
As for circular orbit in equatorial plane, $\dot{r} = 0$, $\implies r = \frac{1}{u} =$ constant. We obtain the specific angular momentum from Eq.(\ref{GE-13}) as, 
\begin{equation} \label{GE-14}
\hat{h}^{2} = \frac{\delta \, M \, (1 - 3 \, \tilde{\omega} e^{4a p^{0}} \, u^{2}) \, e^{- 4 a p^{0}} }{ u - M \, u^{2} ( 3 - 5 \, \tilde{\omega} e^{4a p^{0}} \, u^{2})} . 
\end{equation}
By changing in $r$ variable, and introducing dimensionless parameters as $x = \frac{r}{M}$ \& $\hat{\omega} = \frac{\tilde{\omega}}{M^{2}}$, we obtain, 
\begin{equation} \label{GE-15}
\hat{h}^{2} = \frac{\delta \, x^{2} \, M^{2} (x^{2} - 3 \, \hat{\omega}e^{4a p^{0}}) e^{- 2 a p^{0}}}{(x^{3} - 3 \, x^{2} + 5 \hat{\omega} e^{4a p^{0}})} . 
\end{equation}
Solving for specific energy $\hat{k}$, for circular orbit ($\dot{r} = 0$), from Eq.(\ref{GE-7}), we obtain as,
\begin{equation} \label{GE-16}
- \frac{\delta}{r} \, \hat{M}_{\text{eff}} \, e^{4 a p^{0}} + \frac{\hat{h}^{2}}{2 r^{2}} \, e^{8 a p^{0}} - \frac{\hat{h}^{2}}{r^{3}} \, \hat{M}_{\text{eff}} \, e^{8 a p^{0}} =  \frac{1}{2} \, e^{4 a p^{0}} \, (k^{2} - \delta). 
\end{equation}
Solving the above equations, we obtain the specific energy $\hat{k}$ as,
\begin{equation} \label{GE-17}
\hat{k}^{2} = \frac{\delta \, (x^{3} - 2 \, x^{2} + 2 \, \hat{\omega} e^{4a p^{0}})^{2}}{x^{3} . (x^{3} - 3 \, x^{2} + 5 \, \hat{\omega} e^{4a p^{0}})} . 
\end{equation}
Similarly, for angular velocity $\hat{\Omega} = \dfrac{\dot{\phi}}{\dot{t}}$, substituting value of $\dot{t}$ \& $\dot{\phi}$ from Eq.\eqref{GE-4} \& Eq.\eqref{GE-5}, respectively, we obtain as,
\begin{equation} \label{GE-18}
\hat{\Omega} = \dfrac{ e^{2 a p^{0}} \sqrt{x^{2} - 3 \, \hat{\omega} e^{4a p^{0}}} }{M x^{5/2}} . 
\end{equation}
Substituting the value of $\hat{h}$ from Eq.(\ref{GE-15}) in Eq.(\ref{GE-9}), and re-writing in terms of dimensionless parameter, we obtain the effective potential as,
\begin{equation} \label{GE-19}
\hat{V}_{\text{eff}}^{\text{RGI}} = - \dfrac{ \delta (x^{2} - \, \hat{\omega} e^{4 a p^{0}}) e^{4 a p^{0}}}{x^{3}} . \bigg\{ 1 + \frac{(x^{2} - 3 \, \hat{\omega} e^{4 a p^{0}})}{x^{3} - 3 x^{2} + 5 \hat{\omega} e^{4 a p^{0}}} \bigg\} +   \dfrac{ \delta (x^{2} - 3 \, \hat{\omega} e^{4 a p^{0}}) e^{4 a p^{0}}}{2 (x^{3} - 3 x^{2} + 5 \hat{\omega} e^{4 a p^{0}})} . 
\end{equation}

\begin{figure}[H]
	\centering
	\includegraphics[scale=.55]{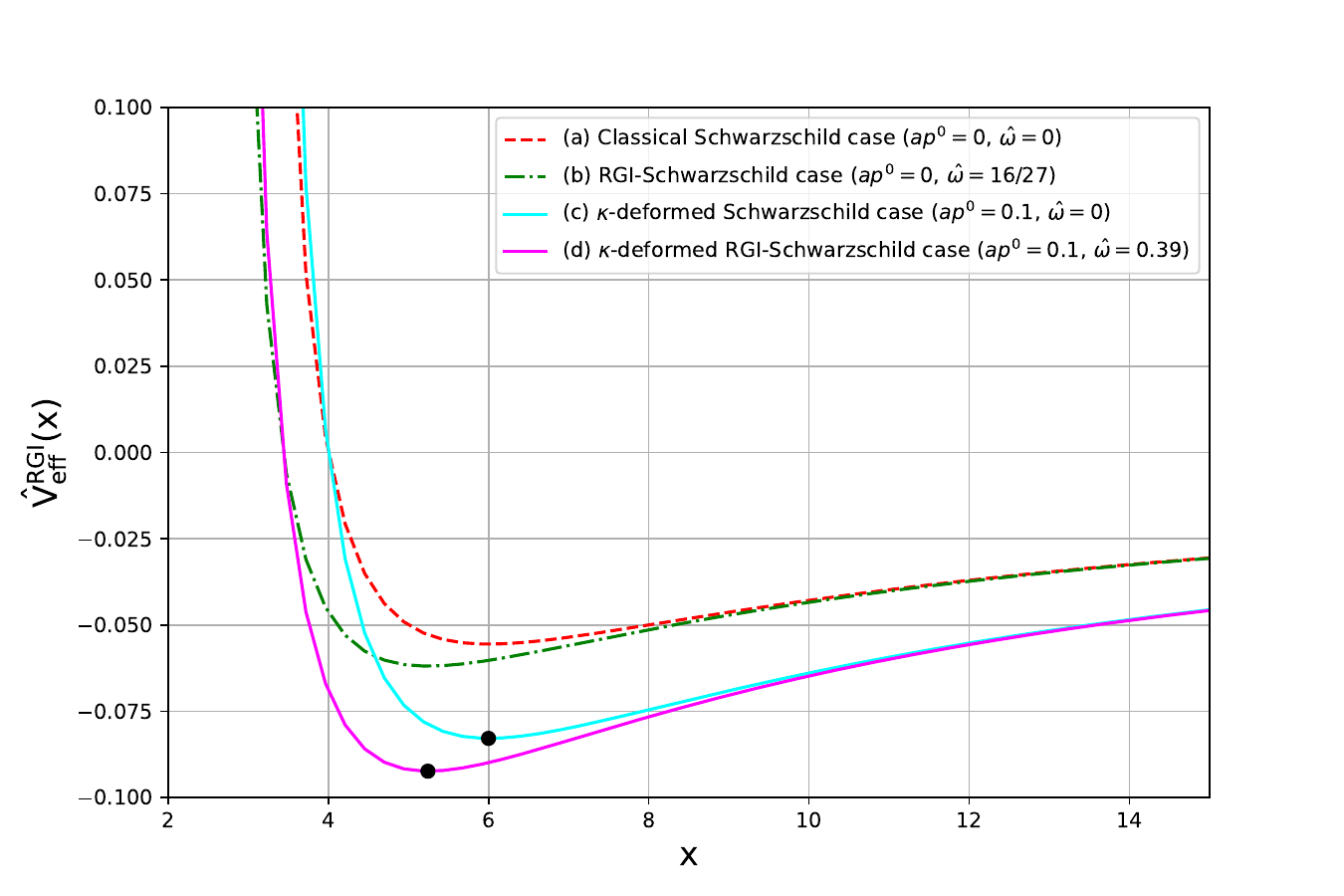}
	\caption{The plot shows the variation of the effective potential $\hat{\text{V}}_{\text{eff}}^{\text{RGI}}$ with respect to the dimensionless radial distance $x=r/M$ for massive particles around Schwarzschild black hole in different space-time backgrounds. The dashed line (red color) represents the classical Schwarzschild case, the dashed-dotted line (green color) corresponds to the RGI-Schwarzschild case in commutative space-time, the solid line (cyan color) shows the $\kappa$-deformed Schwarzschild case, and the solid line (magenta color) represents the $\kappa$-deformed RGI-Schwarzschild case. The $\kappa$-deformation and running gravitational coupling modify the depth and location of the potential well, leading to noticeable shifts in the stability and position of circular orbits, particularly the ISCO, as shown by black circles.}
	\label{Fig3-Veff}
\end{figure}

We have plotted the effective potential $\hat{V}_{\text{eff}}^{\text{RGI}}$ from Eq.(\ref{GE-19}) for Schwarzschild black hole in different space-time background as shown in Fig.(\ref{Fig3-Veff}). We observe from the plot that quantum gravity corrections introduced through the running parameter $\tilde{\omega}$ and non-commutative effect through the deformation parameter $a$ significantly modify the shape of the effective potential compared to the classical Schwarzschild case. In the classical Schwarzschild case, the potential has a broader and shallower well, indicating a weaker gravitational pull near the black hole. When the running of gravitational coupling in commutative space-time is included (RGI-Schwarzschild case), the potential well becomes deeper, showing that gravity becomes stronger at short distances. Adding the $\kappa$-deformation enhances this effect, especially when combined with the running coupling, as seen in the $\kappa$-deformed RGI-Schwarzschild case. The potential well becomes noticeably deeper and narrower, with the peak moving inward. This suggests that the particle experiences a stronger gravitational attraction and can remain in stable orbits much closer to the black hole. These changes shift the location of the minimum of effective potential $V_{\text{eff}}$, meaning the ISCO radius decreases in the modified space-time geometry.

From Fig.(\ref{Fig3-Veff}), we observe that the potential forms a well-shaped curve with a distinct minimum, and this minimum point shifts inward due to the quantum gravity effect. This minimum point corresponds to the innermost stable circular orbit (ISCO). Physically, the ISCO marks the smallest radius at which a particle can maintain a stable circular orbit around the black hole. If a particle moves slightly inward from this point, the orbit becomes unstable, and the particle eventually spirals into the black hole \cite{Kaplane}. Therefore, identifying the position of the minimum in the $V_{\text{eff}}$ curve allows us to determine the ISCO radius. The ISCO radius ($x_{\text{ISCO}}$) is calculated by solving the conditions $\frac{d V_{\text{eff}}}{dx} = 0$ \cite{Page}. Therefore, we solve the equation and we obtain,
\begin{equation} \label{GE-20}
x^{5} + 3 x^{3} \hat{\omega} e^{4 a p^{0}} - 6 x^{4} + 20 x^{2} \hat{\omega} e^{4 a p^{0}} - 30 \hat{\omega}^{2} e^{8 a p^{0}} = 0 . 
\end{equation}
By solving Eq.(\ref{GE-20}), we find that the ISCO radius depends on the running parameter $\hat{\omega}$ and the $\kappa$-deformation parameter $a$. However, the quantum gravity effect from the renormalization group in non-commutative space-time does cause a noticeable shift. Specifically, solving Eq.(\ref{GE-20}) yields $x_{ISCO} = 5.24$ for the $\kappa$-deformed RGI-Schwarzschild case with $\hat{\omega} = 0.39$, while in the classical Schwarzschild case ($\hat{\omega} = 0$), we obtain classical $x_{ISCO} = 6$ \cite{Hobson}. This demonstrates that increasing running parameter $\hat{\omega}$ in non-commutative space-time shifts the ISCO radius inward, highlighting the effect of quantum gravity corrections from the running gravitational coupling in non-commutative space-time geometry.

\begin{figure}[H]
	\centering
	\includegraphics[scale=.55]{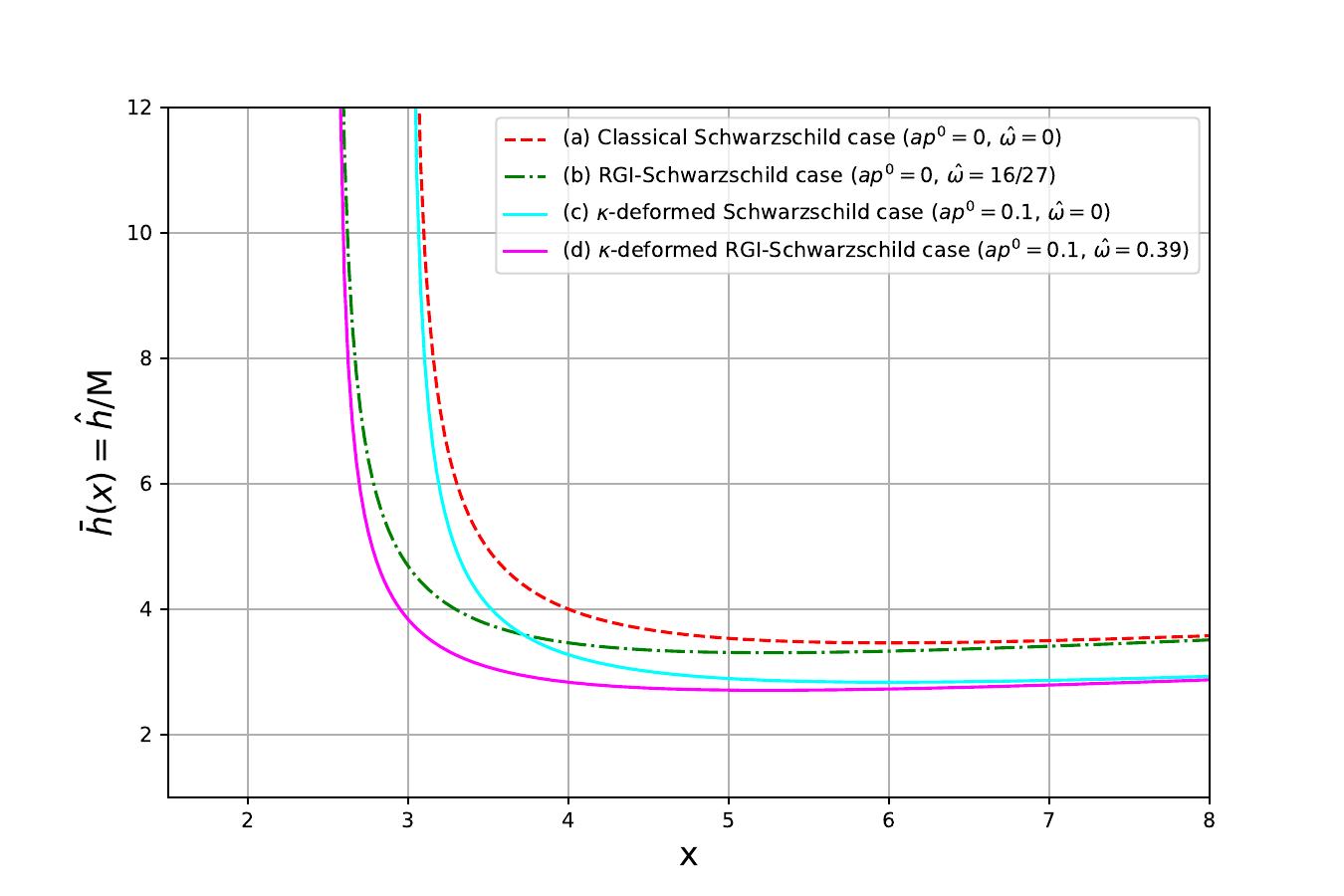}
	\caption{The plot shows the variation of the specific angular momentum $\bar{h} = \hat{h}/M$ with respect to the dimensionless radial distance $x = r/M$ for massive particles around a Schwarzschild black hole in different space-time backgrounds. The dashed line (red color) represents the classical Schwarzschild case, the dashed-dotted line (green color) corresponds to the RGI-Schwarzschild case in commutative space-time, the solid line (cyan color) shows the $\kappa$-deformed Schwarzschild case, and the solid line (magenta color) represents the $\kappa$-deformed RGI-Schwarzschild case.}
	\label{Fig4-angular_momentum}
\end{figure}

To further understand the effect of non-commutativity and RG-induced corrections on particle motion, we now study the behaviour of specific angular momentum. From Eq.(\ref{GE-15}), it is evident that the specific angular momentum of a particle in a circular orbit is influenced by the deformation parameter $a$ and the running parameter $\hat{\omega}$. Specific angular momentum is crucial for understanding orbital dynamics, as it determines the angular motion required to sustain a stable circular orbit at a given radius. Any deviation in angular momentum directly affects the structure of accretion disks and the particle motion near the black hole. Thus, we plotted the specific angular momentum $\bar{h} = \hat{h}/M$ from Eq.(\ref{GE-15}) as shown in the Fig.(\ref{Fig4-angular_momentum}).

From Fig.(\ref{Fig4-angular_momentum}), we observe significant deviations in the behaviour of specific angular momentum $\bar{h}$ across different space-time backgrounds. In the classical Schwarzschild case, the angular momentum increases monotonically with radius, representing the well-known general relativistic profile. When we include the running gravitational coupling through the RGI framework in commutative space-time, the curve shifts downward, indicating that less angular momentum is required to sustain circular orbits at smaller radii. This effect becomes even more pronounced when the RGI correction included in $\kappa$-deformation, where the angular momentum curve shows a sharper decrease near the black hole. These results highlight that the non-commutativity in scale-dependent gravity reduce the required angular momentum for a given orbit, effectively deepening the gravitational potential well. This behaviour is consistent with our findings from the $V_{\text{eff}}$ analysis and supports the conclusion that quantum corrections can significantly alter particle dynamics near the black hole.

\begin{figure}[H]
	\centering
	\includegraphics[scale=.55]{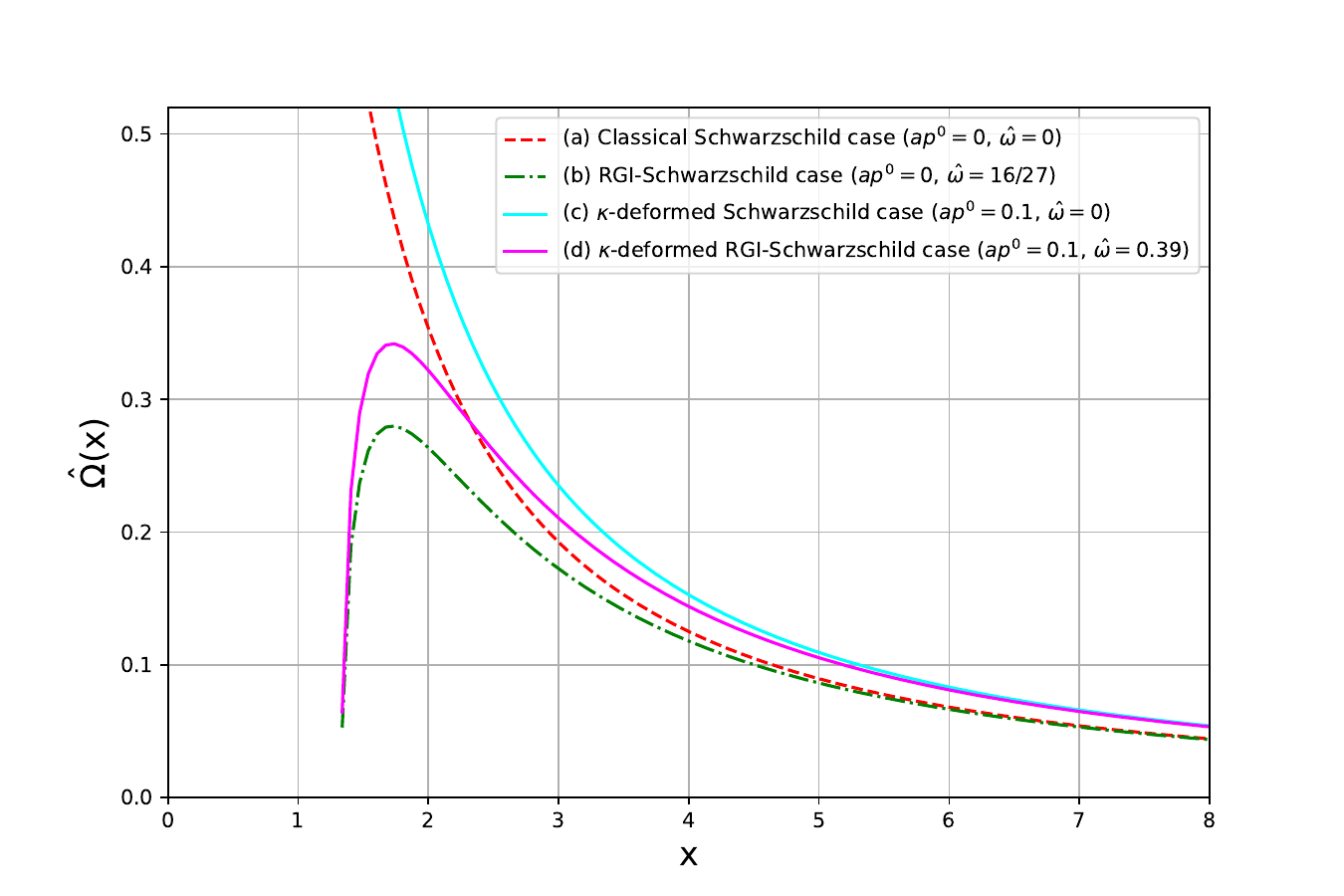}
	\caption{The plot shows the variation of the angular velocity $\hat{\Omega} (x)$ with respect to the dimensionless radial distance $x = r/M$ for massive particles around a Schwarzschild black hole in different space-time backgrounds. The dashed line (red color) represents the classical Schwarzschild case, the dashed-dotted line (green color) corresponds to the RGI-Schwarzschild case in commutative space-time, the solid line (cyan color) shows the $\kappa$-deformed Schwarzschild case, and the solid line (magenta color) represents the $\kappa$-deformed RGI-Schwarzschild case.}
	\label{Fig5-angularVelocity}
\end{figure}

We further investigate the influence of quantum gravity corrections on the angular velocity $\hat{\Omega}$ of test particles in circular orbits. As we obtain from Eq.(\ref{GE-18}), the angular velocity is affected by the non-commutative deformation parameter $a$ and the running parameter $\hat{\omega}$, making it a valid quantity to probe the underlying space-time structure. Angular velocity determines how quickly a particle revolves around the black hole and plays a key role in the dynamics of accretion disks, particularly in governing the rate of energy dissipation and radiation emission. Therefore, we plot the angular velocity $\hat{\Omega}(x)$ from Eq.(\ref{GE-18}), for different space-time, as shown in Fig.(\ref{Fig5-angularVelocity}).

From Fig.(\ref{Fig5-angularVelocity}), we observe notable differences in the angular velocity $\hat{\Omega}(x)$ across various space-time backgrounds. In the classical Schwarzschild case, $\hat{\Omega}(x)$ decreases monotonically with radial distance, consistent with general relativity. When quantum effects are introduced through the RGI framework in commutative space-time, the angular velocity is uniformly suppressed relative to the classical profile. Interestingly, in this case, a small peak develops before the monotonic fall-off, which originates from the scale dependence of the gravitational coupling, showing that quantum corrections weaken gravity at short distances, but $G(r)$ gradually approaches its classical value at large radii, producing a transient enhancement in $\hat{\Omega}(x)$. By contrast, in the purely $\kappa$-deformed Schwarzschild space-time, $\hat{\Omega}(x)$ is enhanced at all radii, suggesting that non-commutativity effectively strengthens the gravitational pull, leading to more compact orbital configurations. When quantum gravitational effects are incorporated through the running gravitational coupling through the RGI framework in non-commutative space-time, the resulting angular velocity again develops a peak. However, its overall profile lies below the purely $\kappa$-deformed case while remaining higher than in the classical Schwarzschild geometry. This highlights the competing influences of scale-dependent gravity and non-commutative deformation on orbital dynamics near the black hole.

\begin{figure}[H]
	\centering
	\includegraphics[scale=.55]{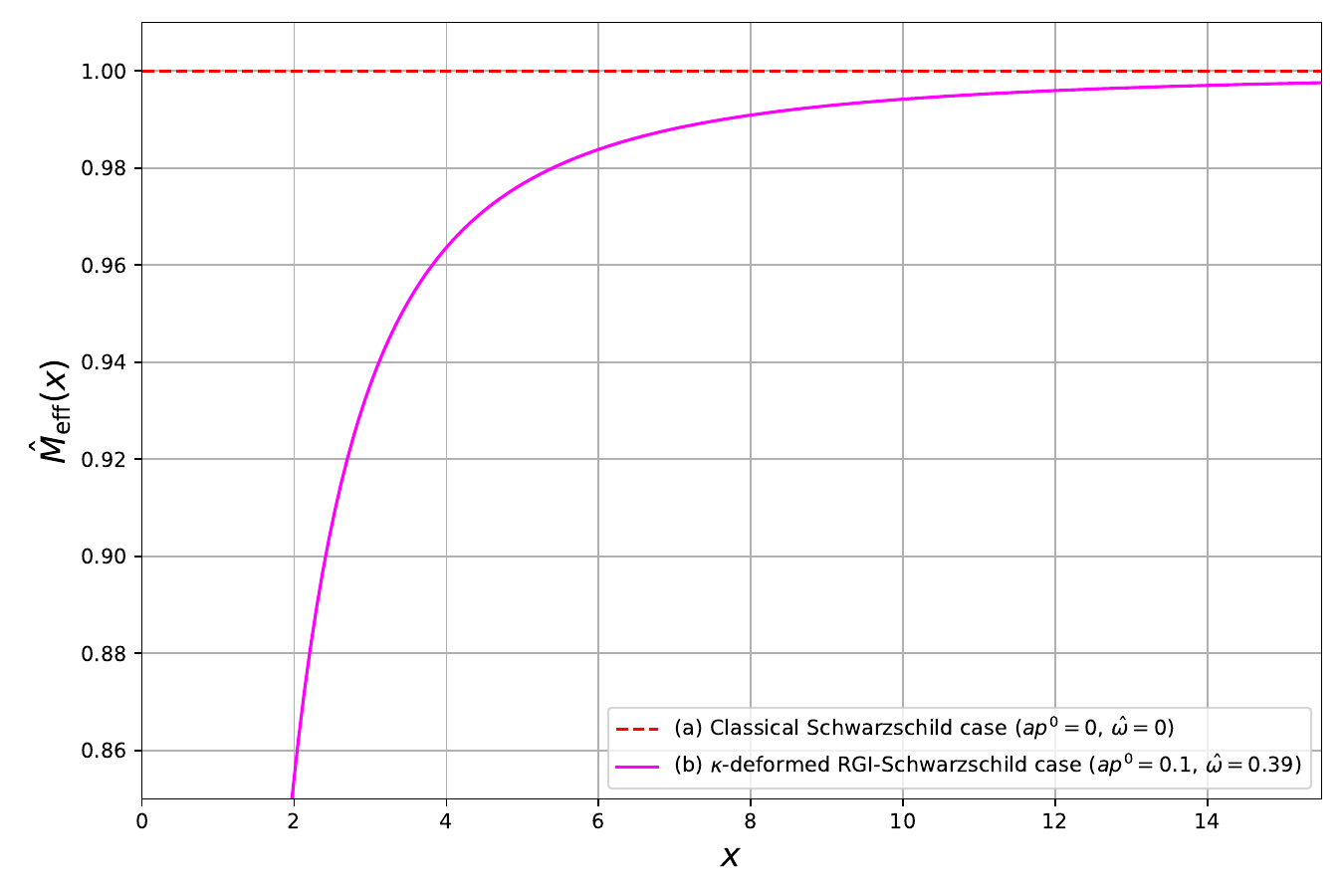}
	\caption{The plot shows the variation of the effective mass $\hat{M}_{\mathrm{eff}}(x)$ with respect to the dimensionless radial distance $x = r/M$ for different space-time backgrounds. The dashed line (red color) represents the classical Schwarzschild case, while the solid line (magenta color) corresponds to the $\kappa$-deformed RGI-Schwarzschild case.}
	\label{Fig-EffectiveMass}
\end{figure}

We further investigate the effect of quantum gravity corrections on the effective mass $M_{\mathrm{eff}}$, which directly governs the strength of the gravitational pull felt by test particles in the vicinity of the black hole. As defined in Eq.(\ref{GE-1}), the effective mass depends on both the running parameter $\tilde{\omega}$ and the non-commutative parameter $a$. The behavior of $\hat{M}_{\mathrm{eff}}(x)$ reveals how the underlying space-time structure modifies the gravitational coupling relative to the classical Schwarzschild case. Therefore, we plot the effective mass $\hat{M}_{\mathrm{eff}}(x)$ from Eq.(\ref{GE-1}) for different space-time geometries, as shown in Fig.(\ref{Fig-EffectiveMass}).

From Fig.(\ref{Fig-EffectiveMass}), we observe significant modifications in the effective mass profile across different space-time geometries. In the classical Schwarzschild case, the mass remains constant, reflecting the absence of quantum or non-commutative effects. In contrast, in the $\kappa$-deformed RGI-Schwarzschild geometry, $\hat{M}_{\mathrm{eff}}(x)$ is reduced at small radii due to the combined influence of scale-dependent gravity and non-commutativity. This suppression indicates that the effective gravitational mass of the black hole weakens in the near-horizon region, modifying the orbital dynamics of particles and the behavior of accretion flows. However, at large $r$, $\hat{M}_{\mathrm{eff}}(x)$ asymptotically approaches $M$, ensuring consistency with the classical limit of general relativity. This analysis demonstrates that $\hat{M}_{\mathrm{eff}}(x)$ provides a clear signature of how quantum corrections and $\kappa$-deformation jointly affect the effective gravitational interaction near the black hole.

\section{Thermal Properties of the Accretion Disk in $\kappa$-deformed RGI-Schwarzschild space-time}				\label{accretion}

When a particle approaches a massive compact object like a black hole, its angular momentum prevents it from falling directly into the black hole. Instead, the particles gradually spiral under gravity, forming a rotating, disk-like structure known as an accretion disk. These disks are among the universe's most luminous and energetic features, radiating large amounts of energy as gravitational potential energy converted into heat. The first study of the foundation for accretion disks used a Newtonian approach \cite{Pringle}. Later, the study advanced using a general relativistic treatment by Page, Thorne, and Novikov \cite{Page,Thorne,Novikov}. Among the various theoretical models of accretion disks, one of the simplest is the steady-state thin disk model \cite{Shakura}. In this model, the disk is assumed to be geometrically thin, allowing the heat produced by internal stresses and friction to be efficiently radiated away from its surface.

The thin accretion disk model considers a disk in a quasi-steady state in the equatorial plane of a stationary, axially symmetric spacetime. The disk material is assumed to move in circular geodesic orbits. The disk is considered thin, implying its half-thickness $h$ is much smaller than its radius $R$ (i.e., $h/R<<1$). The heat generated by internal stresses and friction is efficiently radiated away, primarily from the disk surface. The physical quantities describing the thermal properties of the disk are averaged over the azimuthal angle $\phi = 2\pi$, the vertical height $h$, and a time interval $\Delta t$ corresponding to the gas flowing inward by a distance $2h$ \cite{Zuluaga}. Under these assumptions, the radial structure of the disk is determined by the conservation laws of mass, energy, and angular momentum.

In general relativity, the mass conservation is governed by the continuity equation as \cite{Shakura},
\begin{equation}  \label{T-1}
\nabla_\mu (\rho u^\mu) = 0,
\end{equation}
where $\rho$ is the rest-mass density of accreting fluid, and $u^\mu$ is the four-velocity of the fluid. For a geometrically thin accretion disk confined to the equatorial plane ($\theta = \pi/2$), the four-velocity can be written as,  
\begin{equation} \label{T-2}
u^\mu = \frac{d x^\mu}{d \tau} = (u^{t}, u^{r}, 0, u^{\phi}).
\end{equation}
Thus, the continuity equation Eq.(\ref{T-1}) simplifies for a geometrically thin accretion disk located in the equatorial plane around a black hole in covariant form is given by \cite{Abbas}
\begin{equation} \label{T-3}
\nabla_\mu (\rho u^\mu)  = \frac{1}{\sqrt{-g}} \partial_\mu \left( \sqrt{-g} \, \rho u^\mu \right) = 0,
\end{equation}
here, $\sqrt{-g}$ denotes the determinant of the induced spacetime metric on the equatorial plane at $\theta = \pi/2$. In the steady-state limit, the accretion flow is time-independent, and under the axisymmetry (invariance under rotation about the $z$-axis), all physical quantities depend only on the radial coordinate $r$. Consequently, the continuity equation simplifies to,
\begin{equation} \label{T-4}
\sqrt{-g} \, \rho \, u^{r} = \text{constant},
\end{equation}
which expresses the conservation of mass flux through concentric rings of the disk. To compute the total mass accretion rate, we integrate the mass flux over a ring in the equatorial plane,
\begin{equation} \label{T-5}
\dot{M} = -\int_0^{2\pi} d\phi \int \sqrt{-g} \, \rho \, u^r  \, dz.
\end{equation}
In the thin-disk approximation, the accreting matter is confined to a small vertical extent around the equatorial plane. The vertical integration of the density then defines the surface density as  $ \Sigma(r) = \int \rho \, dz $. Substituting this into Eq.(\ref{T-5}), the total mass accretion rate becomes,
\begin{equation} \label{T-6}
\dot{M} = -\int_0^{2\pi} \Sigma \, u^r \, \sqrt{-g} \, d\phi.
\end{equation}
For classical Schwarzschild space-time, $g_{tt} = - f_{0}(r)$,  $g_{rr} = 1/f_{0}(r)$ and $g_{\phi\phi} = r^{2} \sin^{2}\theta$. Thus, we obtain the determinant of the induced space-time metric for the equitorial plane at $\theta = \pi/2$ as $\sqrt{-g} = r$. Substituting in Eq.(\ref{T-6}), we obtain the total mass accretion rate in classical Schwarzschild case as \cite{Shakura},
\begin{equation} \label{T-7}
\dot{M} = -2\pi r \Sigma \, u^r.
\end{equation}
To compute the total mass accretion rate in the $\kappa$-deformed RGI-Schwarzschild space-time, we must include the effect of non-commutativity. As discussed in section(\ref{geodesic}), $\kappa$-deformed Schwarzschild space-time alters the motion of particles, leading to corrections in the geodesic equations. As a result, the radial velocity ($u^r$) of the accreting matter and the determinant of the space-time metric ($\sqrt{-g}$) are modified. Thus, for the total mass accretion rate in $\kappa$-deformed RGI-Schwarzschild space-time, we integrate the mass flux over a ring in the equatorial plane as,
\begin{equation} \label{T-8}
\dot{\hat{M}} = -\int_0^{2\pi} d\phi \int \sqrt{-\hat{g}} \, \rho \, \hat{u}^r  \, dz.
\end{equation}
As in Eq.(\ref{ksp-15}), we have obtained the $\kappa$-deformed metric components as, $\hat{g}_{00} = g_{00}(\hat{y})$, $\hat{g}_{ij} = g_{ij}(\hat{y})e^{-2ap^{0}} $. Thus, for $\kappa$-deformed RGI-Schwarzschild space-time, $\hat{g}_{tt} = -\hat{f}(r)$, $\hat{g}_{rr} = e^{-4ap^0}/\hat{f}(r)$ and $\hat{g}_{\phi\phi} = r^{2} e^{-4ap^0} \sin^{2}\theta$. We obtain the determinant of the induced space-time metric for the equitorial plane at $\theta = \pi/2$ as $\sqrt{-\hat{g}} = r e^{-4ap^0}$.\\
Similarly, to determine the corresponding form of the $\kappa$-deformed radial velocity, we compare the physical (orthonormal) components of the four-velocity. The physical velocity measured by a local observer is defined through the orthonormal component as \cite{Carroll},
\begin{equation} \label{T-9}
u^{\hat{r}} = \sqrt{g_{rr}} \, u^{r},
\end{equation}
where $u^{r}$ is the coordinate velocity. Since the physical velocity must remain invariant under $\kappa$-deformation, we require that
\begin{equation} \label{T-10}
\hat{u}^{\hat{r}} = u^{\hat{r}}.
\end{equation}
From the relation between physical and coordinate components, the equality of physical velocities in Eq.(\ref{T-10}) leads to
\begin{equation} \label{T-11}
\sqrt{\hat{g}_{rr}} \, \hat{u}^{r} = \sqrt{g_{rr}} \, u^{r}.
\end{equation}
Substituting $\kappa$-deformed metric components $\sqrt{\hat{g}_{rr}} = e^{-a p^{0}} \sqrt{g_{rr}}$ in Eq.(\ref{T-11}), we obtain
\begin{equation} \label{T-12}
\hat{u}^{r} = e^{a p^{0}} \, u^{r}.
\end{equation}
Eq.(\ref{T-12}) represents the $\kappa$-deformed radial velocity, where $u^{r}$ denotes the classical radial velocity. Substituting in Eq.(\ref{T-8}), we obtain the mass accretion rate in $\kappa$-deformed RGI-Schwarzschild space-time as,
\begin{equation} \label{T-13}
\dot{\hat{M}} = -2 \pi r \, e^{-3ap^0} \, \Sigma \, u^r.
\end{equation}
This result implies the $\kappa$-deformation of space-time geometry modifies the accretion rate via the deformation parameter $a$. In the commutative limit $a \to 0$, we recover the mass accretion rate for classical Schwarzschild space-time as shown in Eq.(\ref{T-7}).

\begin{figure}[H]
	\centering
	\includegraphics[scale=.55]{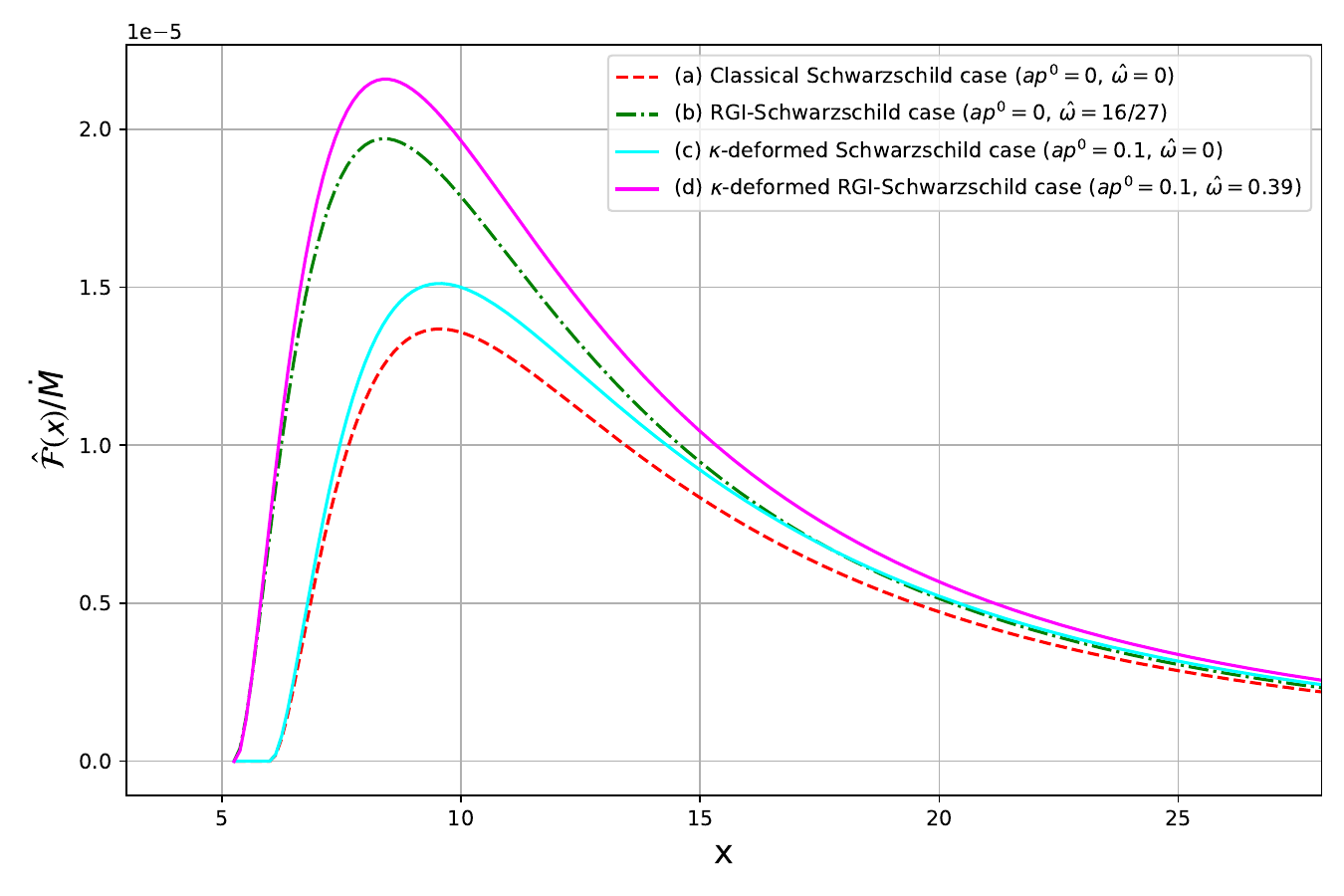}
	\caption{The plot shows the variation of the energy flux $\hat{\mathcal{F}}(x)/\dot{M}$ radiated from the surface of a thin accretion disk around a Schwarzschild black hole in different space-time backgrounds. The dashed line (red color) represents the classical Schwarzschild case, the dashed-dotted line (green color) corresponds to the RGI-Schwarzschild case in commutative space-time, the solid line (cyan color) shows the $\kappa$-deformed Schwarzschild case, and the solid line (magenta color) represents the $\kappa$-deformed RGI-Schwarzschild case. The enhancement of radiated flux in the $\kappa$-deformed RGI-Schwarzschild case reflects stronger gravity near the inner disk region due to non-commutativity and running gravitational coupling.}
	\label{Fig6-Flux}
\end{figure}

In the local rest frame of the accreting fluid, the energy flux emitted from the surface of the disk is defined as the energy radiated per unit area per unit time, given by \cite{Novikov,Joshi}. The radiated energy flux will be modified in the presence of a $\kappa$-deformed RGI-Schwarzschild space-time due to the determinant of the $\kappa$-deformed space-time metric $\sqrt{-\hat{g}}$ and the $\kappa$-deformed mass accretion rate $\dot{\hat{M}}$, $\kappa$-deformation and running gravitational coupling introduces modifications to the values of $k$, $h$, and $\Omega$. We obtain,
\begin{equation} \label{T-14}
\hat{\mathcal{F}}(r) = - \dfrac{\dot{\hat{M}}}{4\pi \, \sqrt{-\hat{g}}} \dfrac{1}{(\hat{k} - \hat{\Omega} \hat{h})^{2}} \bigg(\dfrac{d \hat{\Omega}}{d r}\bigg) \int_{r_{ISCO}}^{r} (\hat{k} - \hat{\Omega} \hat{h}) \, \, \dfrac{d \hat{h}}{d r} \, \, dr. 
\end{equation}
We have plotted the energy flux per unit accretion rate ($\hat{\mathcal{F}}(x)/\dot{M}$), as shown in Fig.(\ref{Fig6-Flux}).   
From Fig.(\ref{Fig6-Flux}), we observe that quantum gravity corrections affect the energy flux radiated from the accretion disk. In the classical Schwarzschild case, the flux profile is shallower and peaks farther from the black hole. Introducing the running gravitational coupling (RGI) or the $\kappa$-deformation causes a noticeable shift in the flux distribution. In particular, the $\kappa$-deformed RGI-Schwarzschild case shows the highest peak and a sharper profile, indicating a stronger gravitational field near the black hole. These results suggest that when both non-commutativity and scale-dependent corrections are present, quantum gravity effects lead to more energetic and compact accretion disks.

Now, we investigate the luminosity of the accretion disks in the $\kappa$-deformed RGI-Schwarzschild space-time as seen by a distant observer. The radial luminosity profile helps us understand how energy is radiated from the different regions of the accretion disk and how these emissions contribute to the total energy output. By combining the conservation laws of energy and angular momentum, we can derive an expression for the differential luminosity observed at infinity as $\mathcal{L}_{\infty}$ \cite{Page, Joshi}. In our study, the space-time geometry and the radiated energy flux are modified due to the effects of $\kappa$-deformation and the running gravitational coupling as the modified metric determinant $\sqrt{-\hat{g}}$, and the modified radiated energy flux $\hat{\mathcal{F}}(r)$ (shown in Eq.(\ref{T-14})). Thus, we obtain the expression for the differential luminosity in the $\kappa$-deformed RGI-Schwarzschild space-time as,
\begin{equation} \label{T-15}
\dfrac{d \hat{\mathcal{L}}_{\infty}}{d \, (\ln r)} = 4\pi \, r \, \sqrt{-\hat{g}} \, \hat{k}  \, \hat{\mathcal{F}}(r). 
\end{equation}

\begin{figure}[H]
	\centering
	\includegraphics[scale=.55]{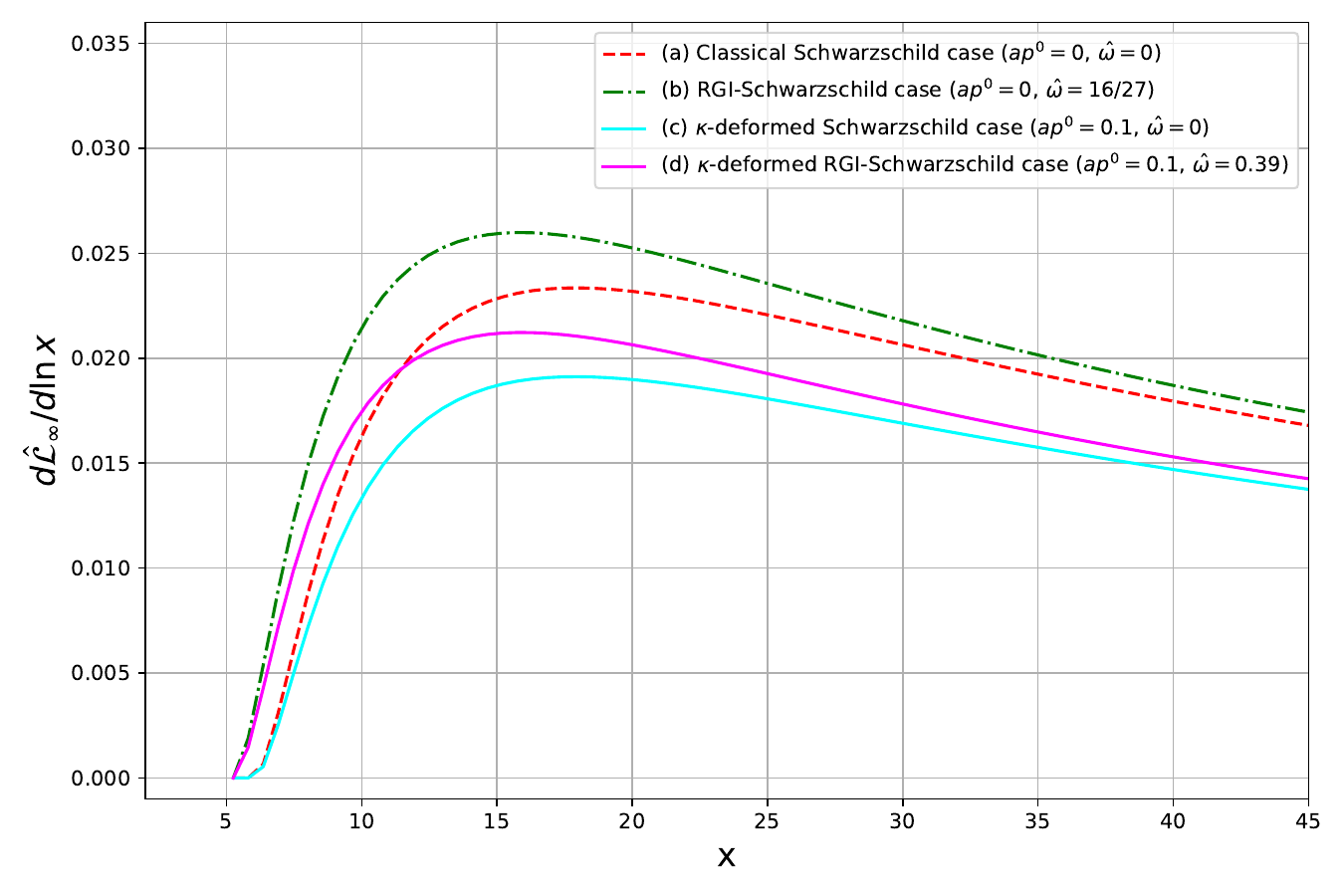}
	\caption{The plot shows the variation of the differential luminosity $d \hat{\mathcal{L}}_{\infty}$ radiated from a thin accretion disk around Schwarzschild black hole in different space-time backgrounds. The dashed line (red color) represents the classical Schwarzschild case, the dashed-dotted line (green color) corresponds to the RGI-Schwarzschild case in commutative space-time, the solid line (cyan color) shows the $\kappa$-deformed Schwarzschild case, and the solid line (magenta color) represents the $\kappa$-deformed RGI-Schwarzschild case.}
	\label{Fig7-Luminosity}
\end{figure}

We have plotted the differential luminosity at infinite ($d \hat{\mathcal{L}}_{\infty} / d \, (\ln r)$) from Eq.(\ref{T-15}), as shown in Fig.(\ref{Fig7-Luminosity}).

From Fig.(\ref{Fig7-Luminosity}), we observe notable differences in the luminosity across the different space-time backgrounds due to the quantum gravity corrections. In the classical Schwarzschild case, the luminosity rises gradually and peaks at a larger radius. For the RGI-Schwarzschild case in commutative space-time, the peak of the luminosity curve shifts slightly inward and becomes maximum and sharper, indicating that more energy is radiated closer to the black hole compare to classical Schwarzschild case. In the $\kappa$-deformed Schwarzschild case, we observe the smallest peak height which implies $\kappa$-deformation reduces the energy output. In the $\kappa$-deformed RGI-Schwarzschild case, the luminosity peak is reduced compared to the RGI-Schwarzschild case in commutative space-time. While the running gravitational coupling increases the luminosity by strengthening the gravitational field near the black hole, adding $\kappa$-deformation counteracts this effect and lowers the peak.

\begin{figure}[H]
	\centering
	\includegraphics[scale=.55]{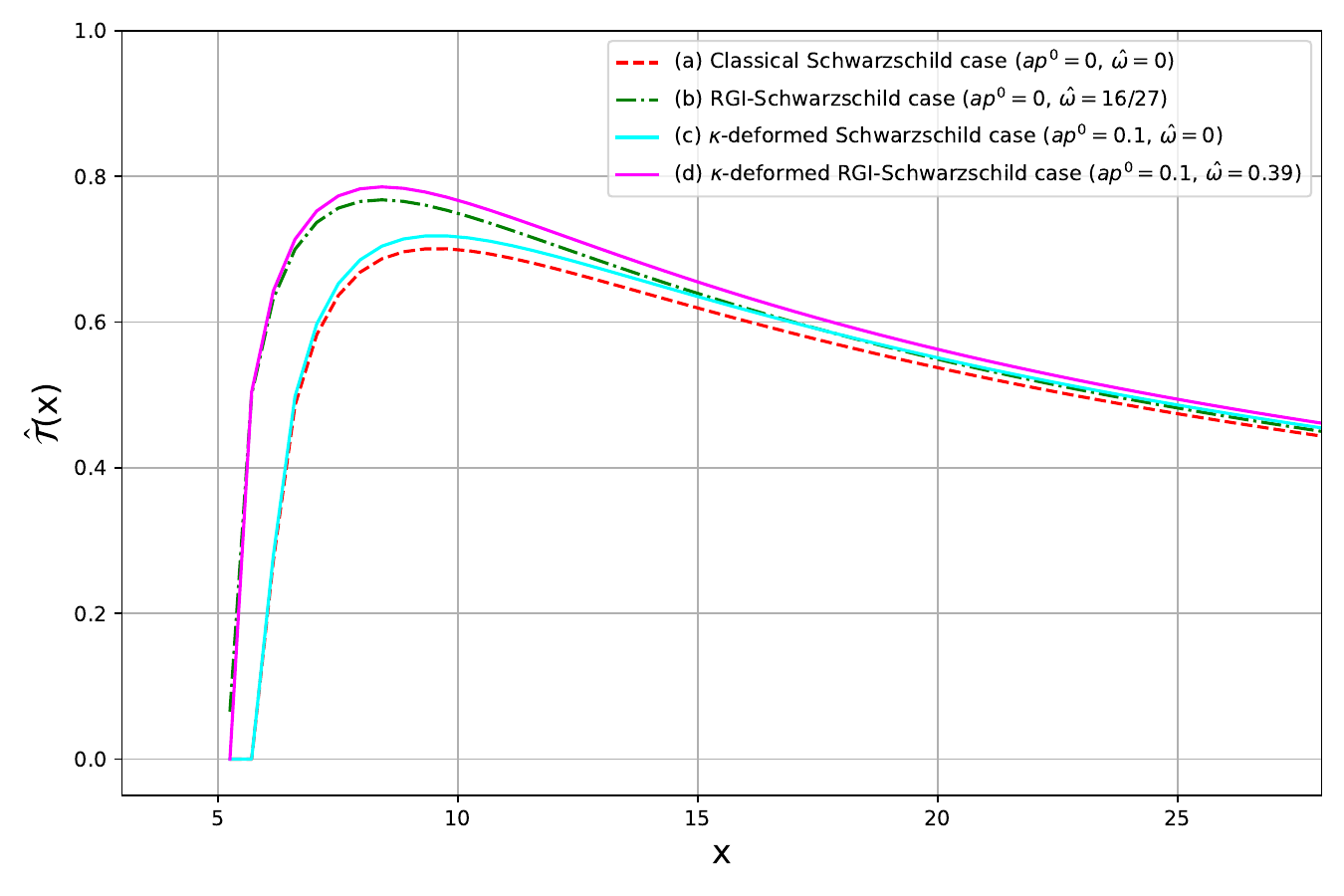}
	\caption{The plot shows the variation of the accretion disk temperature $\hat{\mathcal{T}}(x)$ around a Schwarzschild black hole in different space-time backgrounds. The dashed line (red color) represents the classical Schwarzschild case, the dashed-dotted line (green color) corresponds to the RGI-Schwarzschild case in commutative space-time, the solid line (cyan color) shows the $\kappa$-deformed Schwarzschild case, and the solid line (magenta color) represents the $\kappa$-deformed RGI-Schwarzschild case.}
	\label{Fig8-Temperature}
\end{figure}

We also study the radial profile of the temperature distribution of accretion disks to show the thermal structure and radiative properties of matter near black holes. Since the disk is assumed to be in thermodynamic equilibrium, the radiation emitted from the accretion disk can be considered as blackbody radiation, with the temperature at each radius determined by the local energy flux through the Stefan–Boltzmann law. In our study, the radiated energy flux is modified due to the effects of $\kappa$-deformation and the running gravitational coupling as the modified radiated energy flux $\hat{\mathcal{F}}(r)$ (shown in Eq.(\ref{T-14})). Thus, we obtain the expression for the radial profile of temperature in the $\kappa$-deformed RGI-Schwarzschild space-time as,
\begin{equation} \label{T-16}
\hat{\mathcal{T}}(r) = \sigma^{-1/4} \, \hat{\mathcal{F}}(r)^{1/4}. 
\end{equation} 
where,  $\sigma$ is Stephan Boltzmann constant, i.e. $\sigma = 5.67 \times 10^{-5}$ (in CGS unit.)
We have plotted the radial profile of temperature of accretion disks ($\hat{\mathcal{T}}(x)$) from Eq.(\ref{T-16}), as shown in Fig.(\ref{Fig8-Temperature}).

From Fig.(\ref{Fig8-Temperature}), we observe that the $\kappa$-deformation and running gravitational coupling strongly influence the thermal behaviour of the accretion disk. The classical Schwarzschild background yields the lowest temperature curve while including either RG corrections or $\kappa$-deformation alone leads to a moderate increase in the peak disk temperature. However, the temperature rises more steeply in the $\kappa$-deformed RGI-Schwarzschild spacetime. This is consistent with the behaviour of the energy flux, suggesting that the combined quantum gravity corrections increase the rate of gravitational energy conversion into heat. The sharper and higher peak also implies that more energetic thermal radiation is emitted from the inner regions of the disk in the deformed geometry.

\section{Conclusion}    			\label{conclusion}

In this work, we have investigated the combined effects of non-commutativity and quantum gravitational corrections on the dynamics of test particles and thermal properties of accretion disks around a Schwarzschild black hole. Specifically, we studied particle motion and disk behaviour in the background of a $\kappa$-deformed RGI-Schwarzschild space-time, which incorporates a running gravitational coupling $\hat{G}(r)$ obtained from the renormalization group (RG) approach in non-commutativity characterized by the deformation parameter $a$ arising from $\kappa$-deformed space-time.

We began by starting with the classical Schwarzschild metric and introducing quantum gravitational corrections through a scale-dependent Newton’s constant $\hat{G}(r)$ in  non-commutative space-time, as motivated by the renormalization group (RG) technique. This yielded the $\kappa$-deformed RGI-Schwarzschild metric. The resulting line element captures quantum gravity effects through changes in space-time structure and coordinate algebra. The resulting space-time geometry showed significant deviations near the black hole horizon. By analyzing the horizon structure, we identified a critical value of the RG parameter, $\tilde{\omega}_c = 0.39$, beyond which the black hole no longer has an event horizon. This indicates a transition from a black hole to a naked singularity.

Using this modified metric, we derived the geodesic equations for a test particle and examined the influence of both the running coupling and non-commutative effect on particle trajectories. Through numerical analysis, we demonstrated how the combined effects of $a$ and $\tilde{\omega}$ alter the orbits of particles, showing noticeable deviations from classical and RGI-Schwarzschild backgrounds in commutative space-time. The trajectories became more tightly bound, indicating enhanced gravitational attraction due to the quantum-corrected space-time.

Further, we analyzed the effective potential in the $\kappa$-deformed RGI-Schwarzschild background and computed the innermost stable circular orbit (ISCO) radius. The effective potential curves showed that non-commutativity and RG-improvement deepen the potential well, making particle orbits more tightly confined and leading to a reduction in the ISCO radius from $x = 6$ (in classical Schwarzschild) to $x \simeq 5.24$ for $\tilde{\omega} = 0.39$. The ISCO radius affected by the RGI-Schwarzschild space-time in non-commutative space-time geometry.

In addition to the effective potential and ISCO radius, we investigated the behaviour of specific angular momentum, specific energy, and angular velocity for particles in circular orbits. Our analysis revealed that the specific angular momentum, specific energy, and angular velocity are influenced by the running parameter $\tilde{\omega}$ and the $\kappa$-deformation, indicating that these quantities are affected by both quantum gravity corrections. We illustrated these results by plotting the specific angular momentum and angular velocity, showing how quantum corrections modify the dynamical properties of particles moving around the black hole.

We extend our study by evaluating key thermal properties of the accretion disk, such as the radiation flux, temperature profile, and differential luminosity. We derived the modified expressions for the radiated flux, differential luminosity, and disk temperature, considering the modified metric coefficient and modified mass accretion rate. The results show that the non-commutative deformation and the running coupling enhance these quantities compared to their classical Schwarzschild background. The increase in thermal output near the inner disk region can be interpreted as a consequence of stronger gravitational attraction induced by quantum gravity corrections, which cause matter to fall in more rapidly and heat up more intensely.

Overall, our findings of the combined influence of non-commutative geometry and renormalization group improvement introduce potentially observable modifications to the physics of accretion around black holes. These deviations might manifest in the emission spectra of accretion disks and could provide indirect evidence of quantum gravitational effects in astrophysical environments. Future extensions of this work could include a generalization to realistic rotating black hole environments, such as accretion disks around an RGI-Kerr black hole, or a comparison with observational data from X-ray binaries and active galactic nuclei. This study not only advances our understanding of black hole accretion in the presence of quantum gravity corrections but also highlights the importance of combining different theoretical frameworks, such as non-commutative geometry and RG-improvement, to gain a more complete picture of gravity at the Planck scale.

\begin{center}
 Acknowledgments
\end{center}  
The computational part of this research work was carried out on the computational facility set up from the funds given by the DST- SERB/ANRF Power Grant no. SPG/2021/002228 of the Government of India. D.K acknowledges financial support from DST- SERB/ANRF Power Grant no. SPG/2021/002228 of the Government of India. D. K. would like to thank Prof. Soma Sanyal and Prof. E. Hari Kumar for their insightful discussions and valuable support in understanding the various technical aspects of this research work. 

\section*{Data availability}

Data sharing is not applicable to this article as no data sets were generated or analyzed during the current study.

\section*{Conflict of interest} 

The authors declare no conflict of interest.

\end{document}